%% file: elektra_ADAM.tex
\documentclass[structabstract]{aa}
\usepackage{amsmath}
\usepackage{graphicx,subcaption,caption}
\usepackage{txfonts, longtable, dcolumn,verbatim}
\usepackage[T1]{fontenc}
\usepackage{natbib}
\bibpunct{(}{)}{;}{a}{}{,}

\usepackage[squaren,textstyle]{SIunits}

\newcommand{\tiu}{\joule\usk\power{\metre}{-2}\power{\second}{-1/2}\power{\kelvin}{-1}}

\begin{document}

   \title{Shape model of asteroid (130)~Elektra from optical photometry and disk-resolved images from VLT/SPHERE and Nirc2/Keck}

   \author{
	      J.~Hanu{\v s}\inst{1,2,3*}
           \and
	      F.~Marchis\inst{4}
           \and           
          M.~Viikinkoski\inst{5}
           \and
          B.~Yang\inst{6}
           \and           
          M.~Kaasalainen\inst{5}
}

   \institute{
	     Centre National d'\'Etudes Spatiales, 2 place Maurice Quentin, 75039 Paris cedex 01, France\\
	     $^*$\email{hanus.home@gmail.com}
	 \and
	     Universit\' e C\^ ote d'Azur, OCA, CNRS, Lagrange, France
	 \and
	     Astronomical Institute, Faculty of Mathematics and Physics, Charles University in Prague, V~Hole{\v s}ovi{\v c}k{\'a}ch 2, 18000 Prague, Czech Republic
 	 \and
	     SETI Institute, Carl Sagan Center, 189 Bernado Avenue, Mountain View CA 94043, USA
         \and
	     Department of Mathematics, Tampere University of Technology, PO Box 553, 33101, Tampere, Finland
	 \and
	     European Southern Observatory (ESO), Alonso de Cordova 3107, 1900 Casilla Vitacura, Santiago, Chile
}

   \date{Received x-x-2016 / Accepted x-x-2016}
 
  \abstract
   {Asteroid (130)~Elektra belongs to one of the six known triple asteroids in the main belt, so its mass has been reliably determined.}
   {We aim to use all available disk-resolved images of (130)~Elektra obtained by the SPHERE instrument at VLT and by the Nirc2 of the Keck telescope together with the disk-integrated photometry to determine its shape model and its size. The volume can be then used in combination with the known mass to derive the bulk density of the primary.}
   {We apply the All-Data Asteroid Modeling (ADAM) algorithm to the optical disk-integrated data, 2 disk-resolved images obtained by the SPHERE instrument and 13 disk-resolved images from the Nirc2 of the Keck telescope, and derive the shape model and size of Elektra.}
   {We present the shape model, volume-equivalent diameter (199$\pm$7 km) and bulk density (1.60$\pm$0.13 g\,cm$^{-3}$) of the C-type asteroid Elektra.}
   {}
 

  \titlerunning{Shape model and size of Elektra}
  \maketitle

\section{Introduction}\label{sec:introduction}


The asteroid (130)~Elektra (hereafter simply Elektra) has been classified as a G-type asteroid in the Tholen system \citep{Tholen1989b} and Ch according to the SMASS II classification \citep{Bus2002}. Elektra is associated with CM chondrites due to the presence of an absorption near 0.7~$\mu$m \citep{Cloutis2012b}.

The binary nature of Elektra was revealed by \citet{Merline2003b} using the Keck-II adaptive optics (AO) system in August 2003 and later confirmed by \citet{Marchis2006}. The second satellite was reported in the images obtained by the Spectro-Polarimetric High-contrast Exoplanet Research instrument (SPHERE) by \citet{Yang2015,Yang2016}. SPHERE is an extreme adaptive optics system and coronographic facility installed at the UT3 Nasmyth focus of the ESO's 8.2-m Very Large Telescope (VLT) \citep{Beuzit2008}. The smaller moon is about 2 km across, and orbiting on an eccentric orbit about 500 km away from the primary. This made Elektra the sixth triple system detected in the asteroid belt (after (45)~Eugenia, (87)~Sylvia, (93)~Minerva, (107)~Camilla, and (216)~Kleopatra). The orbit of Elektra's larger satellite is slightly eccentric (e$\sim$0.1), probably due to tidal excitation. Both moonlets of Elektra orbit well-inside the Hill sphere of the primary. \citet{Yang2016} found that the origin of the moonlets is consistent with a sub-disruptive impact scenario rather than having been captured.

The mass was determined by \citet{Marchis2008a} from the analysis of the moon orbit -- (6.6$\pm$0.4) 10$^{18}$ kg.

Simple shape models, based on rotating ellipsoids, amplitude-aspect or magnitude-aspect, estimate the latitude of Elektra's spin axis to be $\sim$--85$^{\circ}$ in the ecliptic coordinate frame \citep{Drummond1988b, Magnusson1990, Drummond1991, Michalowski1993, DeAngelis1995}. The reported ecliptic longitude of the pole varies significantly within the various solutions, however, they represent almost the same solution. This is because the longitudes are very dense for latitude values close to $\pm$90$^{\circ}$, so even a small distance of two points on a surface results in a large difference in their ecliptic longitudes. For example, our third ADAM solution in Tab.~\ref{tab:spins} ($\lambda$, $\beta$ = 71, $-$88) differs from the \citet{Durech2011} value ($\lambda$, $\beta$ = 64, $-$88) by 7 degrees of longitude. However, in this region so close to the ecliptic pole, 7 degrees of longitude, equates to only 14 minutes of arc. The lightcurve inversion technique \citep{Kaasalainen2001a, Kaasalainen2001b} confirmed the previous pole determinations \citep{Durech2007a, Torppa2008, Hanus2016a}.

Size estimates based on comparison of shape models with disk-resolved data (from Keck) or occultation silhouettes vary between 180 and 215 km \citep{Marchis2006, Marchis2008a, Durech2011, Hanus2013b}. The radiometric sizes based on IRAS, AKARI and WISE data are consistent with this range. However, we do not consider them reliable, because the radiometric method is affected, among others, by the systematic effect of the single epoch observation (i.e., one geometry of observation). Note that the lightcurve amplitude is $\sim$0.4, so this systematic effect could be important. In addition, \citet{Marchis2012b} analyzed Spitzer spectra in mid-IR by the means of a thermophysical model and estimated the size ($D$=197$\pm$20~km), geometric visible albedo ($p_{\mathrm{V}}$=0.064$\pm$0.013) and thermal inertia (5--65 \tiu). 

The density of Elektra was previously determined by \citet{Marchis2012b} (1.7$\pm$0.3 g\,cm$^{-3}$) and by \citet{Hanus2013b} (1.99$\pm$0.66 g\,cm$^{-3}$). The rather large uncertainties are caused by the large uncertainty in the size estimates.

Recently, models combining both disk-integrated and disk-resolved data were developed \citep[e.g., KOALA and ADAM models,][]{Carry2012b,Viikinkoski2015}. With those inversion algorithms, both asteroid's shape and size are derived simultaneously \citep[e.g., asteroids (234)~Barbara or (3)~Juno,][]{Tanga2015,Viikinkoski2015b}. We used the All-Data Asteroid Modeling (ADAM) algorithm here to determine the shape and size of Elektra.

The angular resolution of the SPHERE IFS instrument at the observed wavelength is 0.037'', so a slightly better value compared to the one achieved by the Nirc2 camera on Keck II (0.045''). Combined with the fact that all but one images from Keck II were obtained when the Earth-Elektra distance was larger than for the SPHERE images, the SPHERE images should significantly improve the shape and size estimates for Elektra, so consequently its density. Note that the accurate mass of Elektra can be derived using the well-known orbits of the satellites, so the main uncertainty in the bulk density comes from the size estimate.

In Sec.~\ref{sec:data}, we present optical disk-integrated data, together with the disk-resolved data obtained by the Keck II and VLT/UT3 telescopes equipped with the adaptive optics systems (Nirc2 and SPHERE/IFS). The ADAM algorithm used for the shape and size optimization is described in Sec.~\ref{sec:ADAM}. We present the shape model of Elektra in Sec.~\ref{sec:models} and discuss its physical properties in Sec.~\ref{sec:densities}. Finally, we conclude our work in Sec.~\ref{sec:conclusions}.

\section{Data}\label{sec:data}

\subsection{Optical disk-integrated photometry and convex shape model}\label{sec:photometry}

It is important to have good initial knowledge of the spin period and the spin axis orientation of Elektra. 

An up-to-date convex shape model derived by the lightcurve inversion method \citep{Kaasalainen2001a, Kaasalainen2001b} was recently presented by \citet{Hanus2016a} and made available in the Database of Asteroid Models from Inversion Techniques \citep[DAMIT\footnote{\texttt{http://astro.troja.mff.cuni.cz/projects/asteroids3D}},][]{Durech2010}. To be complete, we list all previous spin axis and shape model determinations in Tab.~\ref{tab:spins}. 

From DAMIT, we downloaded 54 disk-integrated optical lightcurves of Elektra from 13 apparitions (listed in Tab.~\ref{tab:photometry}). The images obtained in standard filter systems were bias- and flat-field corrected using standard procedures. These lightcurves are based on aperture differential photometry using several nearby stars. Although some of the data were initially absolutely calibrated, we used all lightcurves in a relative sense only, meaning that we normalized all of them. For the lightcurve inversion, only the relative change of the brightness due to rotation, shape and orientation with respect to the Sun and the observer is important. The otherwise unknown size can be then constrained by the disk-resolved data. We did not consider the sparse-in-time measurements from astrometric surveys \citep[see, e.g.,][]{Hanus2011} because of their redundancy -- the dense dataset of much higher quality was sufficient for the shape modeling and the sparse data were mostly adding noise.

\subsection{Disk-resolved images}\label{sec:keckAO}

The W.M. Keck II telescope is located at Maunakea in Hawaii. The telescope is equipped since 2000 with an AO system and the near-infrared camera (Nirc2). This AO system provides an angular resolution close to the diffraction limit of the telescope at $\sim$2.2 $\mu$m, so $\sim$45 mas for bright targets (V$<$13.5) \citep{Wizinowich2000}. The AO system was improved several times since it was mounted. For example, the correction quality of the system was improved in 2007 \citep{vanDam2004}, resulting into reaching an angular resolution of 33 mas at shorter wavelengths ($\sim$1.6 $\mu$m).

All data obtained by the Nirc2 extending back to 2001 are available at the Keck Observatory Archive (KOA). It is possible to download the raw images with all necessary calibration and reduction files, and often also images on which basic reduction was performed. We downloaded and processed all disk-resolved images of Elektra. Usually, several frames were obtained by shift-adding 3--17 frames with an exposure time of several seconds depending on the asteroid's brightness at particular epoch. We performed the flat-field correction and used a bad-pixel suppressing algorithm to improve the quality of the images before shift-adding them. Finally, we deconvolved each image by the AIDA algorithm \citep{Hom2007} to improve its sharpness.

Moreover, we also included three images of Elektra already used in our previous work \citep{Marchis2006,Hanus2013b}. These data were processed by a similar pipeline as the data from KOA. A total number of 13 Keck disk-resolved images from 5 different apparitions were obtained, see Tab.~\ref{tab:ao} for additional information.

Our two SPHERE disk-resolved images from December 9 and 30, 2014 (see Tab.~\ref{tab:ao}) were obtained by the Internal Field Spectrograph \citep[IFS,][]{Claudi2008} instrument that allows a spacial resolution of 7.4 mas \citep{Mesa2015}. We observed in the field stabilized mode, where the sky remained fixed with respect to the detector. The fields of view of IFS is 1.73''$\times$1.73'', while the pixel scale is 0.0123''.

We processed the data with the SPHERE consortium's pipeline \citep{Pavlov2008}, which consists of standard procedures such as dark subtraction, bad pixel treatment, flat fielding and wavelength calibration. Next, the data were re-sampled into a cube of 39 images of 3.3\% band width ($\Delta\lambda$/$\lambda$) over the spectral range and with a scale of 0.0074'' per spaxel. 

The disk-resolved SPHERE images were already processed and used by \citet{Yang2016}, however, the authors only focused on the positions of the two satellites of Elektra and did not pay attention to the resolved primary.

A complete list of 15 disk-resolved images is provided in Tab.~\ref{tab:ao}.

\section{Method: All-Data Asteroid Modeling (ADAM) algorithm}\label{sec:ADAM}

All-Data Asteroid Modelling algorithm \citep{Viikinkoski2015, Viikinkoski2016} is a universal inversion technique capable of dealing with various disk-resolved data types (adaptive optics, interferometry, and range-Doppler radar data). Moreover, resolved data can be combined with disk-integrated data (photometry), stellar occultation timings, and thermal infrared data. ADAM minimizes the difference between the Fourier transformed image and a projected polyhedral model. This approach facilitates the usage of adaptive optics images directly, without requiring the extraction of boundary contours.  

More specifically, we minimize an objective function 
\[
\begin{split}
&\sum_i\sum_{j=1}^{N_i}\left\Vert V(u_{ij},v_{ij})-e^{2\pi\imath\left(o^x_i u_{ij}+o^y_i v_{ij}\right)+s_i}\,\mathcal{F}M_i(u_{ij},v_{ij})\right\Vert^2\\&+\chi^2_{LC}+\sum_i\lambda_i\gamma_i^2=:\chi^2,\end{split}
\]
where $V(u_{ij},v_{ij})$ is the Fourier transform of the image and $\mathcal{F}M_i(u_{ij},v_{ij})$ that of the plane-projected model $M$ evaluated at the $j$th frequency point $(u_{ij},v_{ij})$ of the $i$th image. The offset $(o^x,o^y)$ within the plane and the scale $s_i$ are free parameters determined during the optimization. The term $\chi^2_{LC}$ is a square norm measuring the model fit to the lightcurves. The last term corresponds to regularization functions $\gamma_i$ and their weights $\lambda_i$ \citep{Viikinkoski2015}. 

The usage of different shape supports \citep[i.e., subdivision surfaces and octanoids, see][]{Viikinkoski2015} and regularization functions (we penalize large planar surfaces as well as local and global concavities) allows features caused by parametric representations to be distinguished from those supported by the data. In particular, features actually present in the data should be visible in all the shape models with identical $\chi^2$-fits.

\section{Results and discussions}\label{sec:results}

\subsection{Shape model and volume of Elektra}\label{sec:models}

\begin{figure}
  \centering
   \begin{subfigure}[b]{0.33\columnwidth}
      \includegraphics[clip=true,trim=90 100 90 100,scale=0.2]{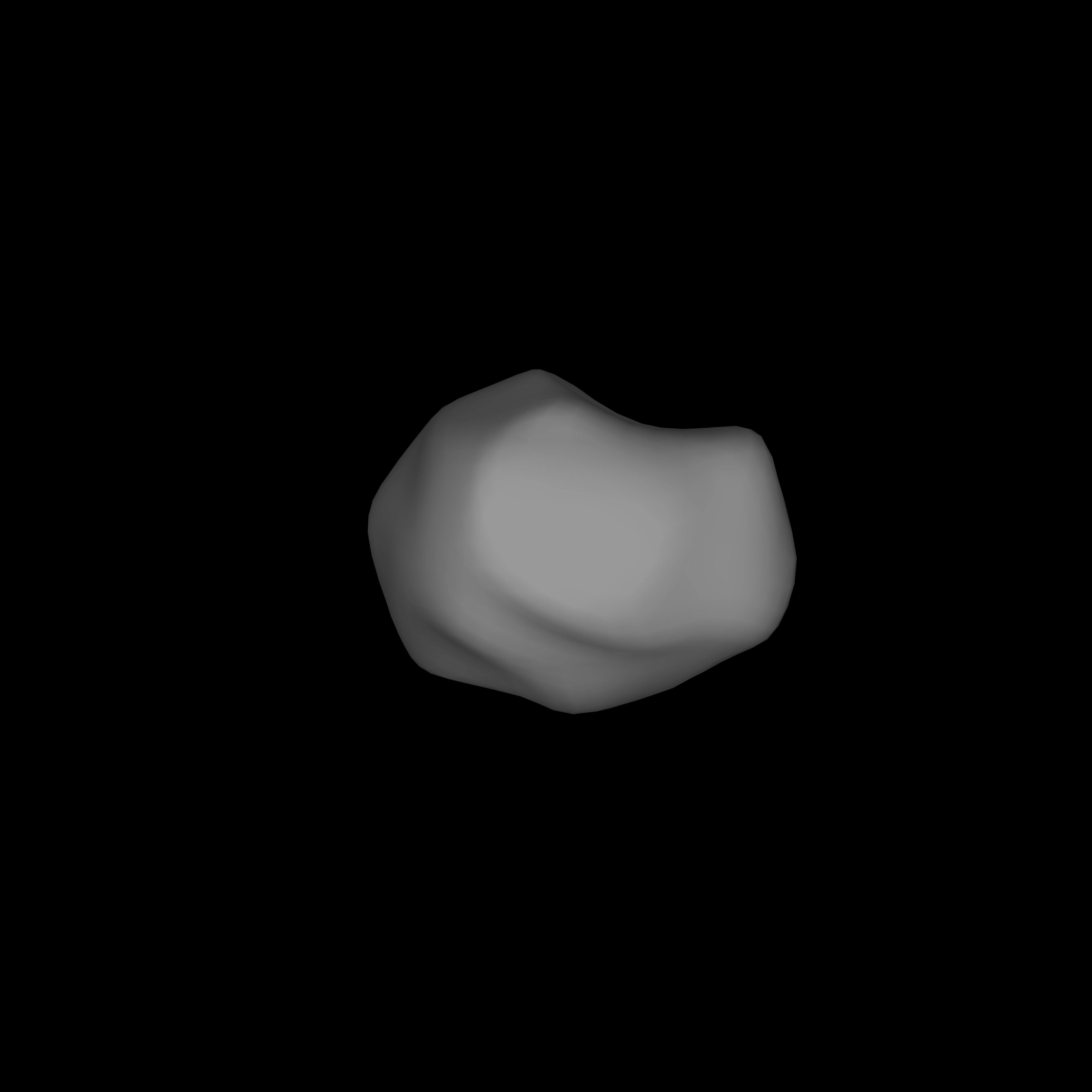}
    \end{subfigure}%
    \begin{subfigure}[b]{0.33\columnwidth}
      \includegraphics[clip=true,trim=90 100 90 100,scale=0.2]{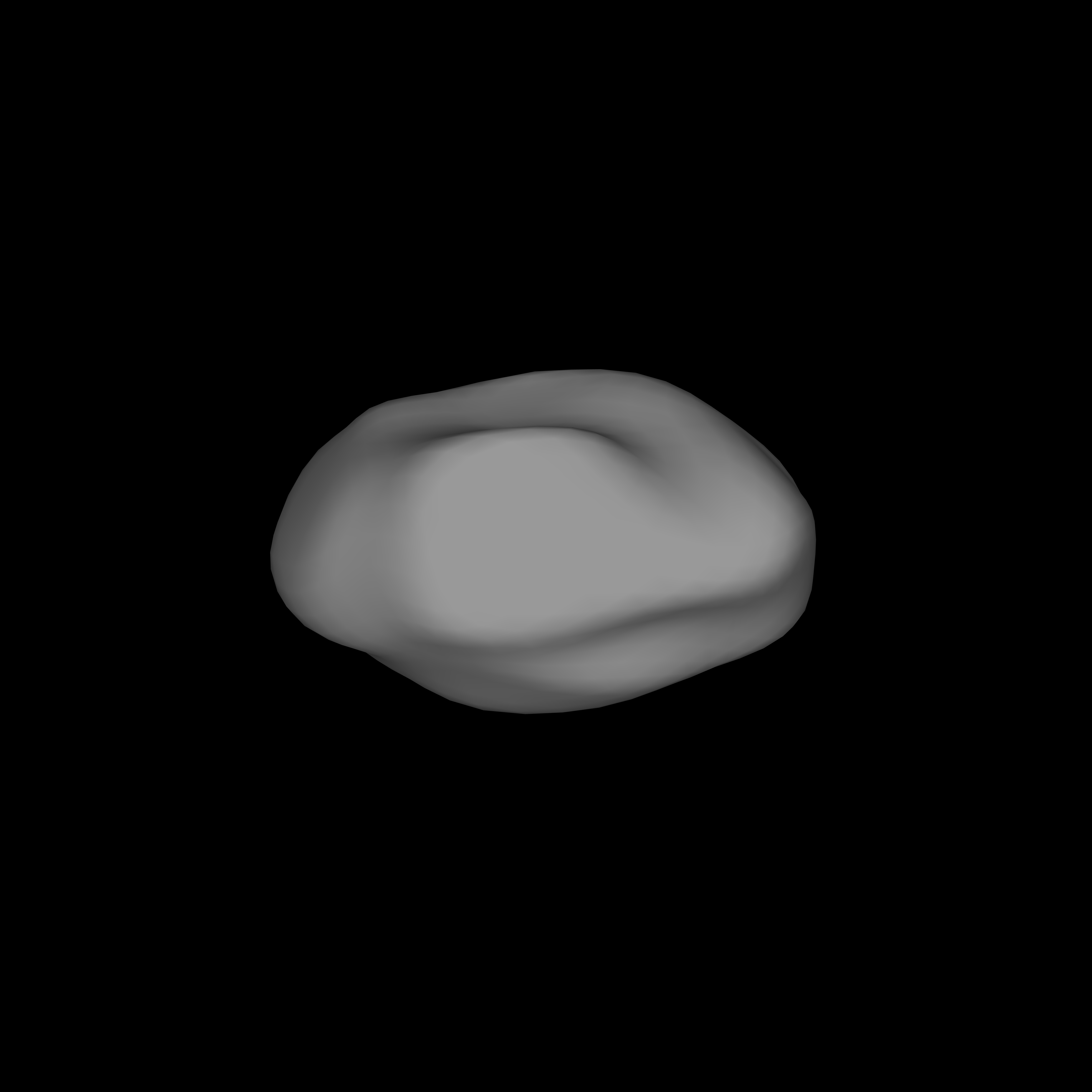}
    \end{subfigure}%
    \begin{subfigure}[b]{0.33\columnwidth}
      \includegraphics[clip=true,trim=90 100 90 100,scale=0.2]{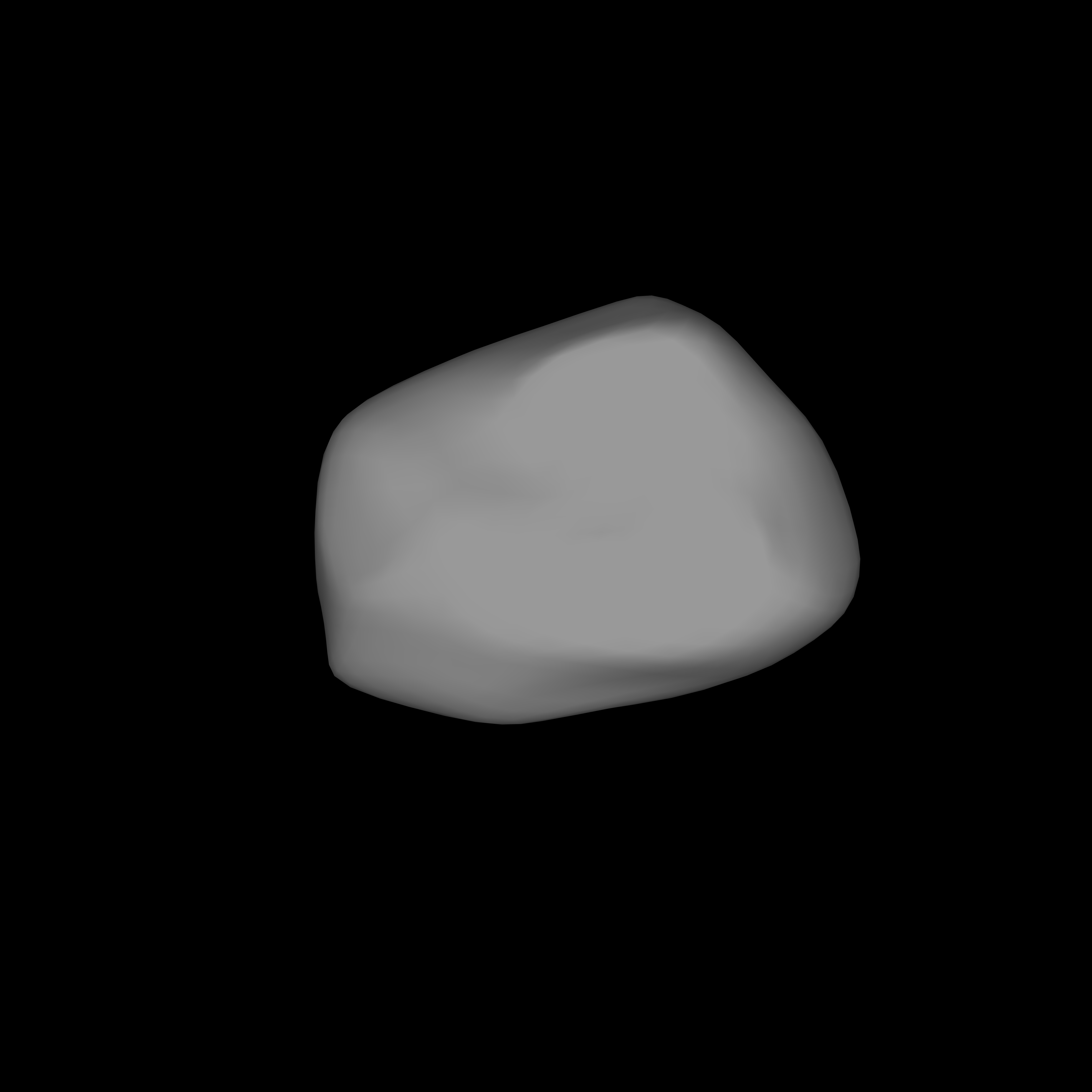}
    \end{subfigure}%
    
    \begin{subfigure}[b]{0.33\columnwidth}
      \includegraphics[clip=true,trim=90 100 90 100,scale=0.2]{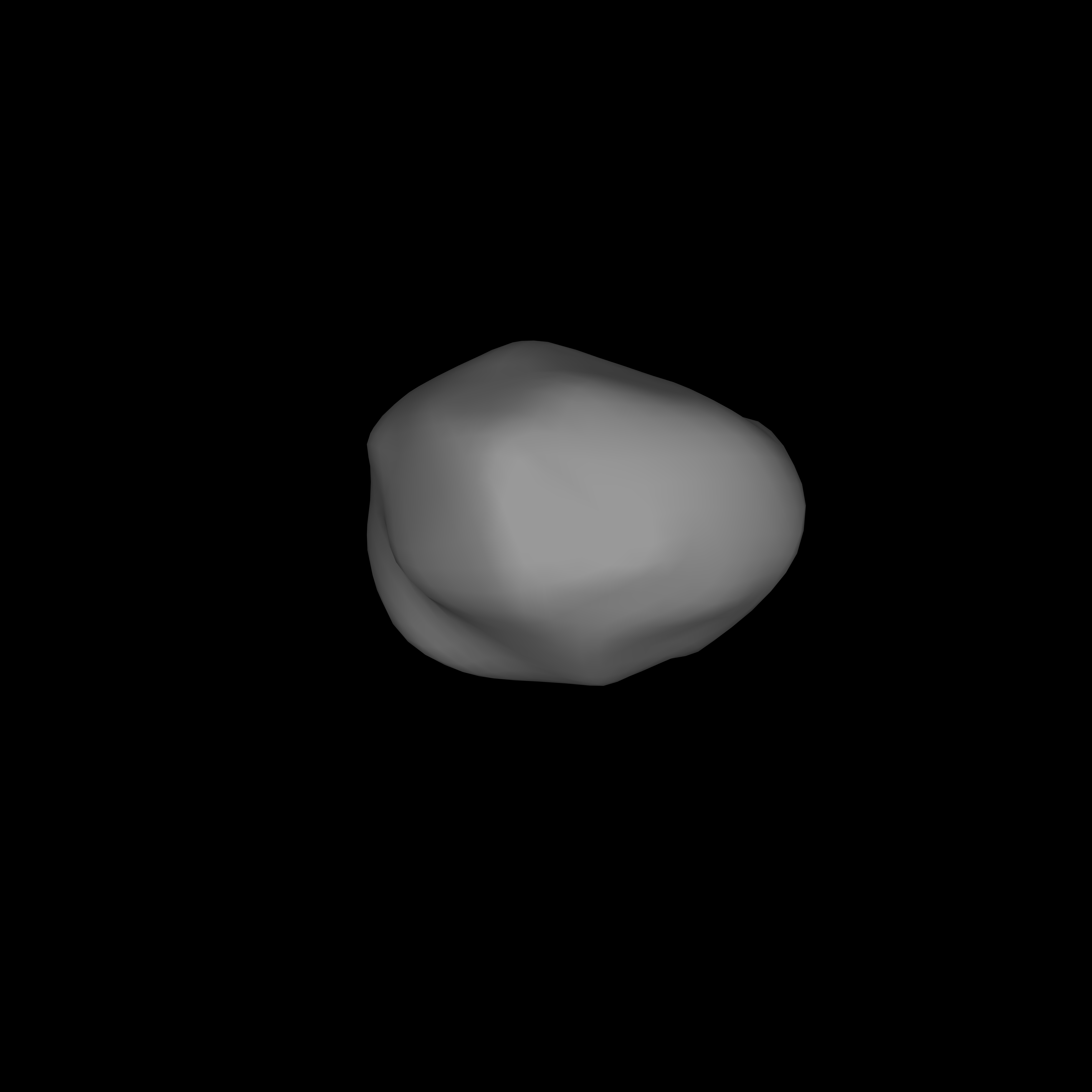}
    \end{subfigure}%
    \begin{subfigure}[b]{0.33\columnwidth}
      \includegraphics[clip=true,trim=90 100 90 100,scale=0.2]{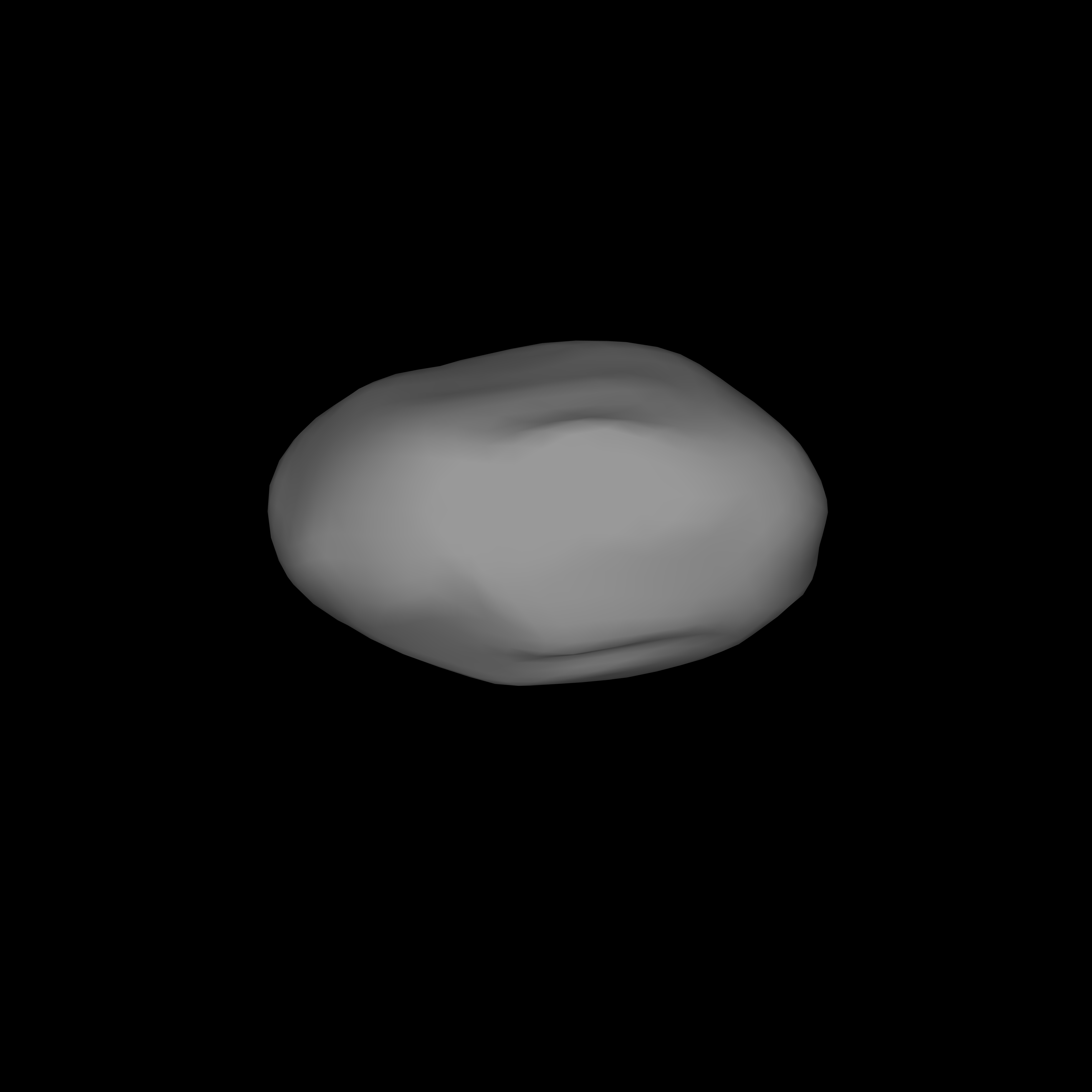}
    \end{subfigure}%
    \begin{subfigure}[b]{0.33\columnwidth}
      \includegraphics[clip=true,trim=90 100 90 100,scale=0.2]{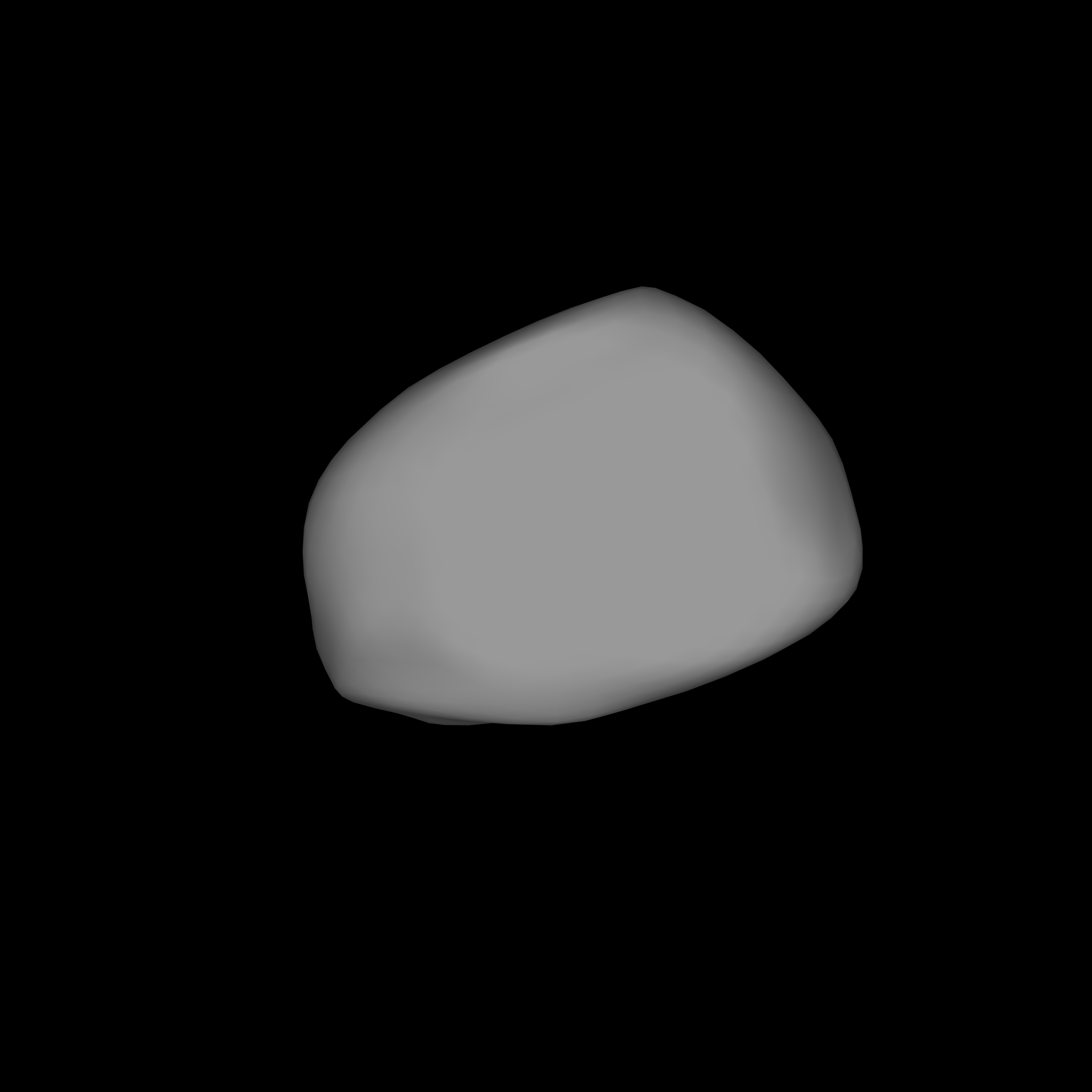}
    \end{subfigure}%
    
    \begin{subfigure}[b]{0.33\columnwidth}
      \includegraphics[clip=true,trim=90 100 90 100,scale=0.2]{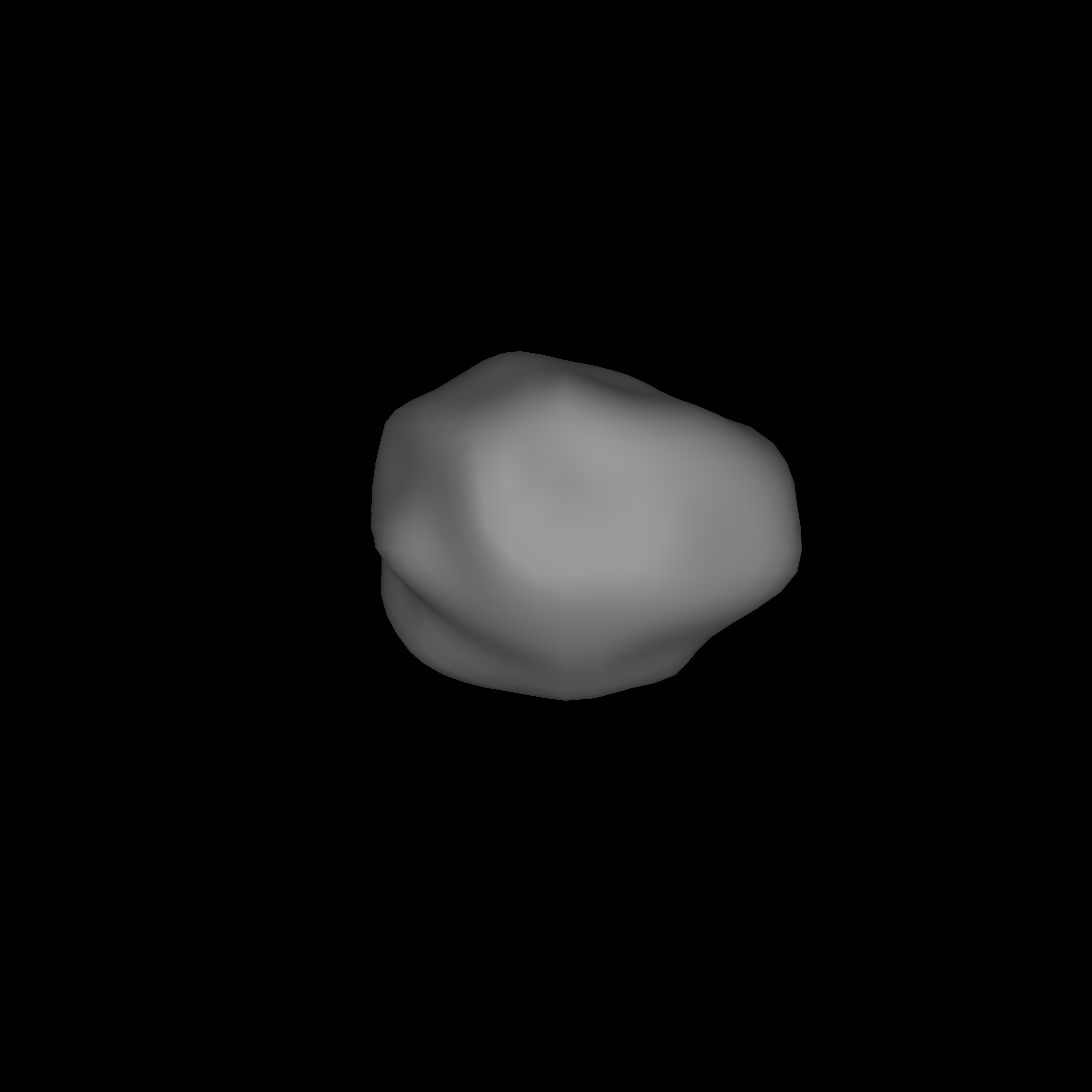}
    \end{subfigure}%
    \begin{subfigure}[b]{0.33\columnwidth}
      \includegraphics[clip=true,trim=90 100 90 100,scale=0.2]{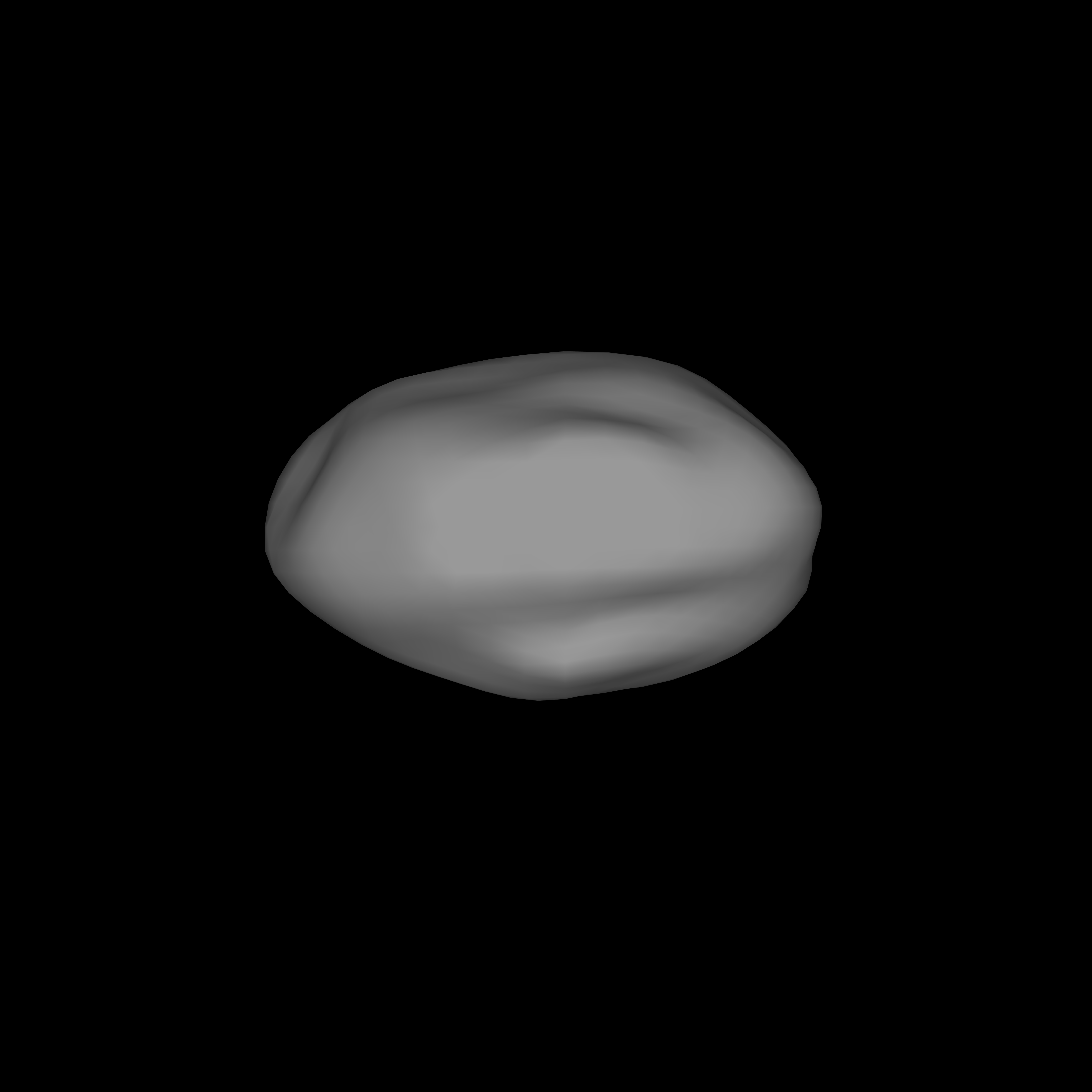}
    \end{subfigure}%
    \begin{subfigure}[b]{0.33\columnwidth}
      \includegraphics[clip=true,trim=90 100 90 100,scale=0.2]{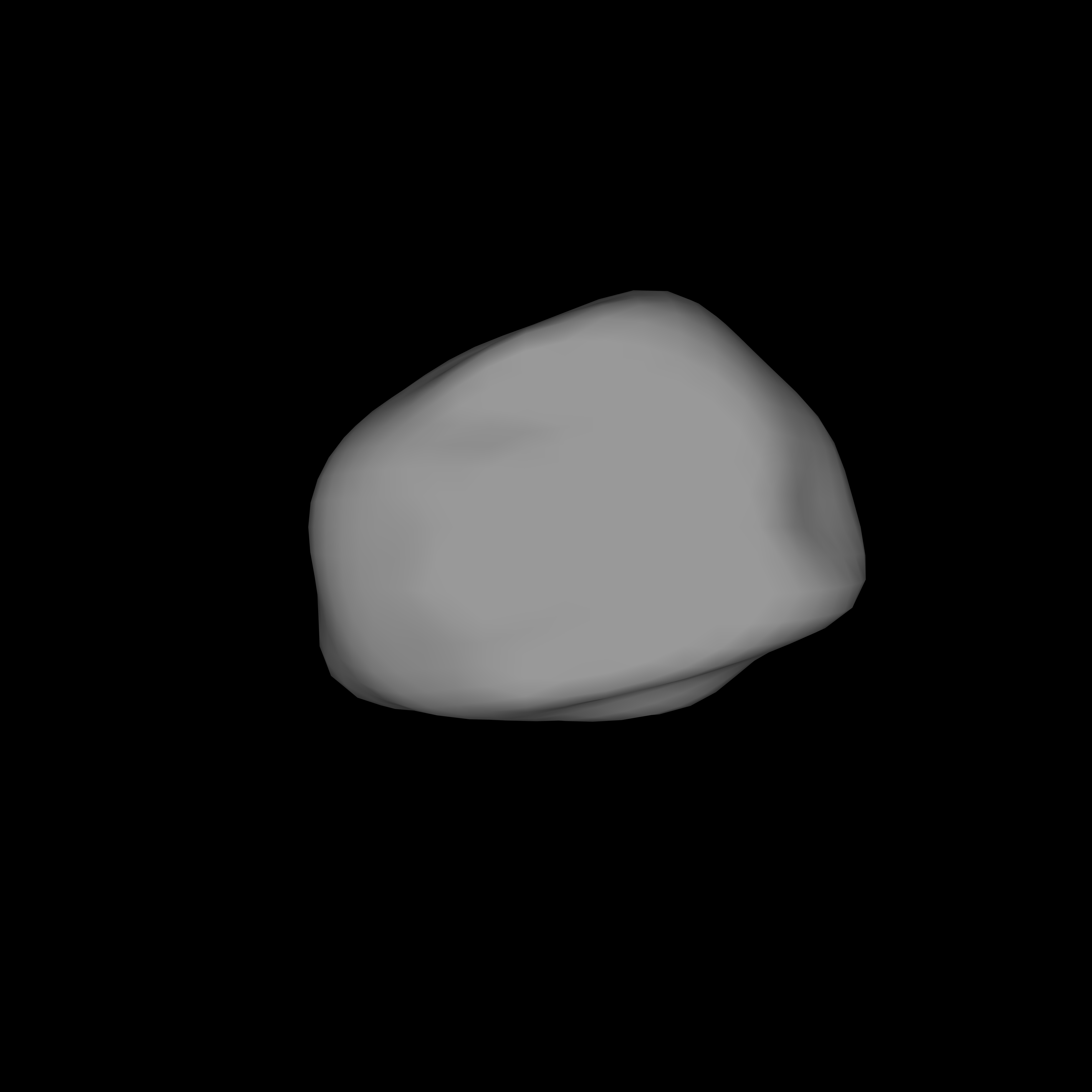}
    \end{subfigure}%
     \caption{\label{fig:shape}Shape model of Elektra reconstructed from disk-integrated optical data and (i)~raw SPHERE images (top panel), (ii)~all resolved images using subdivision surfaces shape support (middle panel), and finally (iii)~all resolved data using octanoids shape support (bottom panel). Each panel shows the shape model at three different viewing geometries: the first two are equator-on views rotated by 90$^{\circ}$, the third one is a pole-on view.}
\end{figure}

\begin{figure*}
  \centering
    \begin{subfigure}[b]{0.247\columnwidth}
      \includegraphics[clip=true,trim=90 100 90 100,scale=0.5]{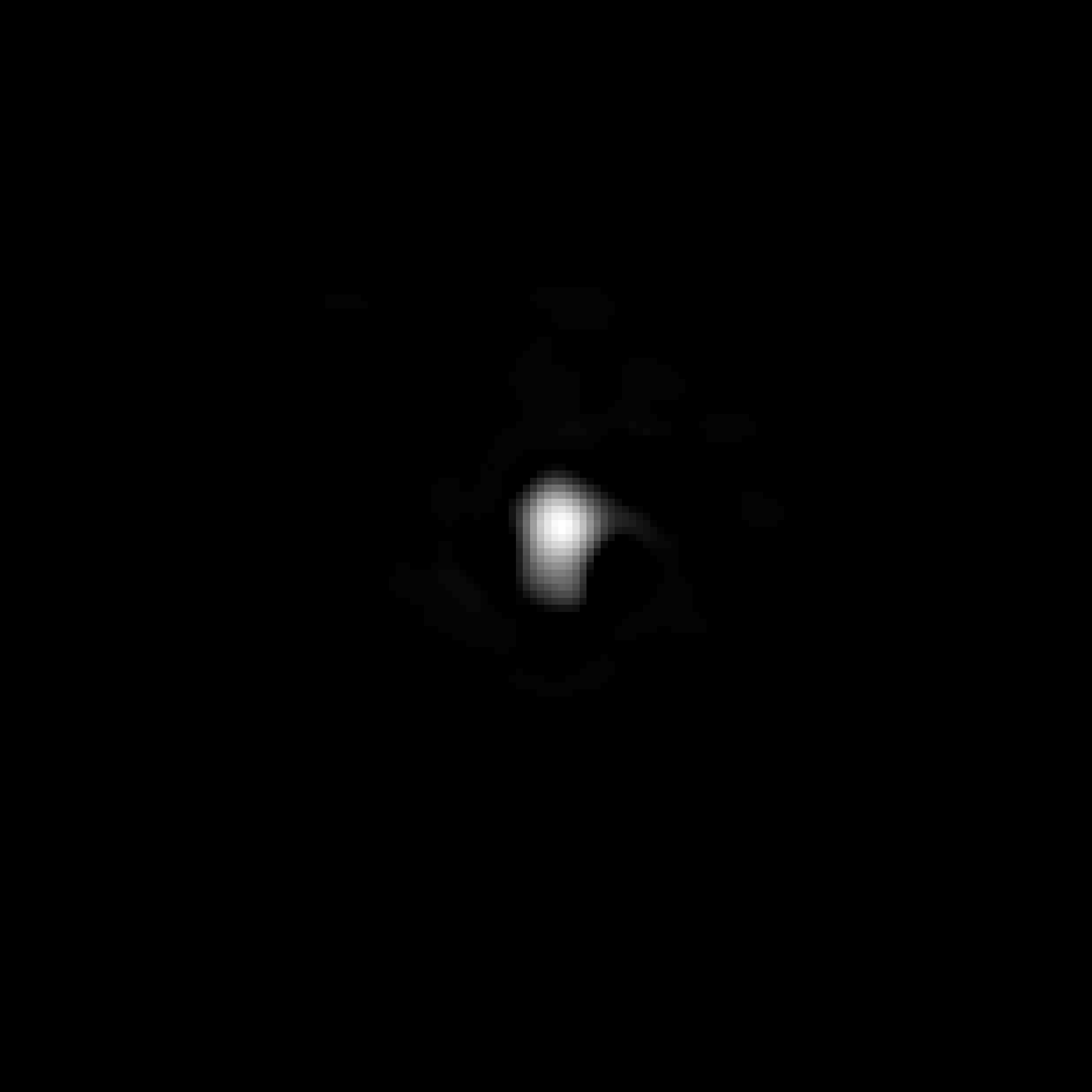}
    \end{subfigure}%
    \begin{subfigure}[b]{0.247\columnwidth}
      \includegraphics[clip=true,trim=90 100 90 100,scale=0.5]{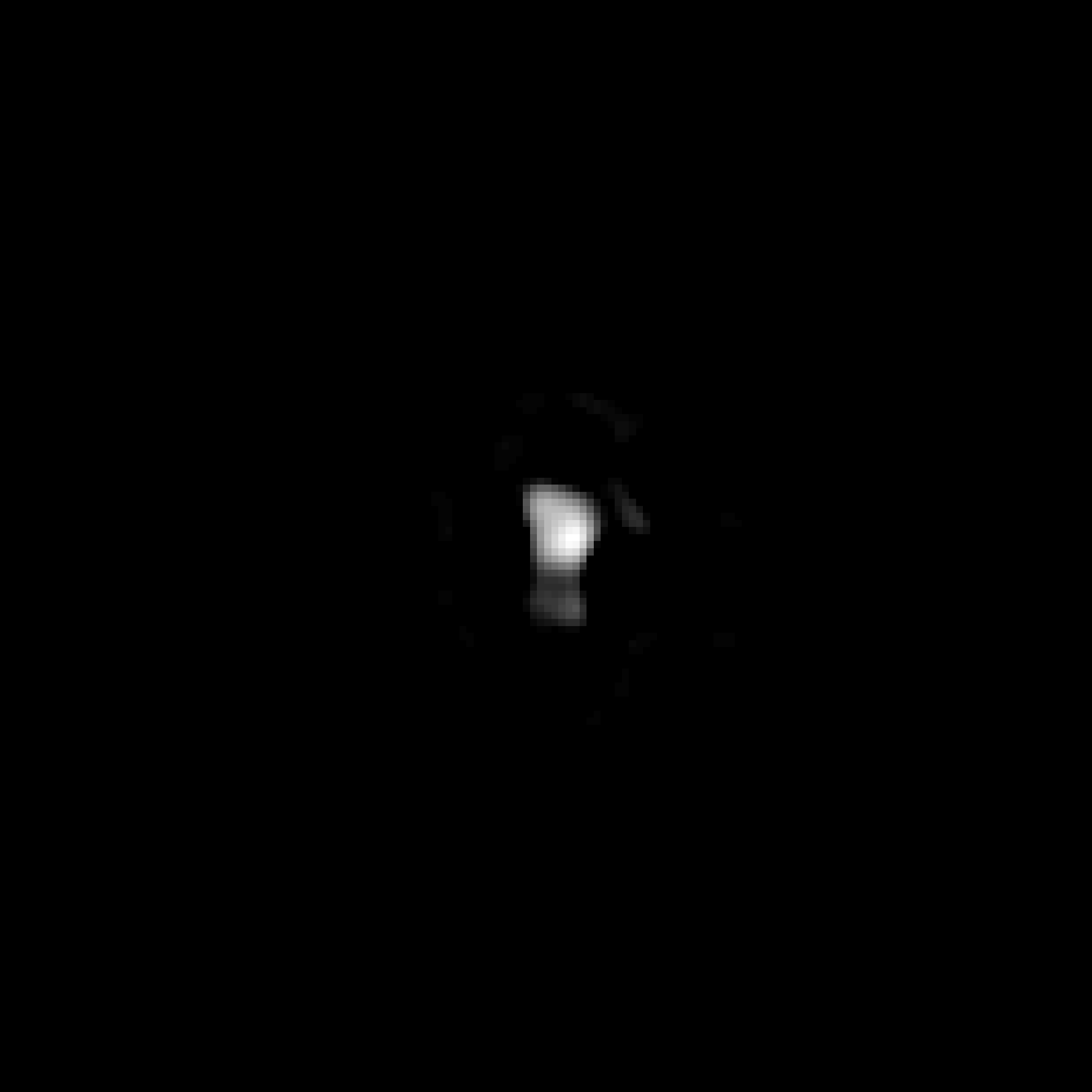}
    \end{subfigure}%
    \begin{subfigure}[b]{0.247\columnwidth}
      \includegraphics[clip=true,trim=90 100 90 100,scale=0.5]{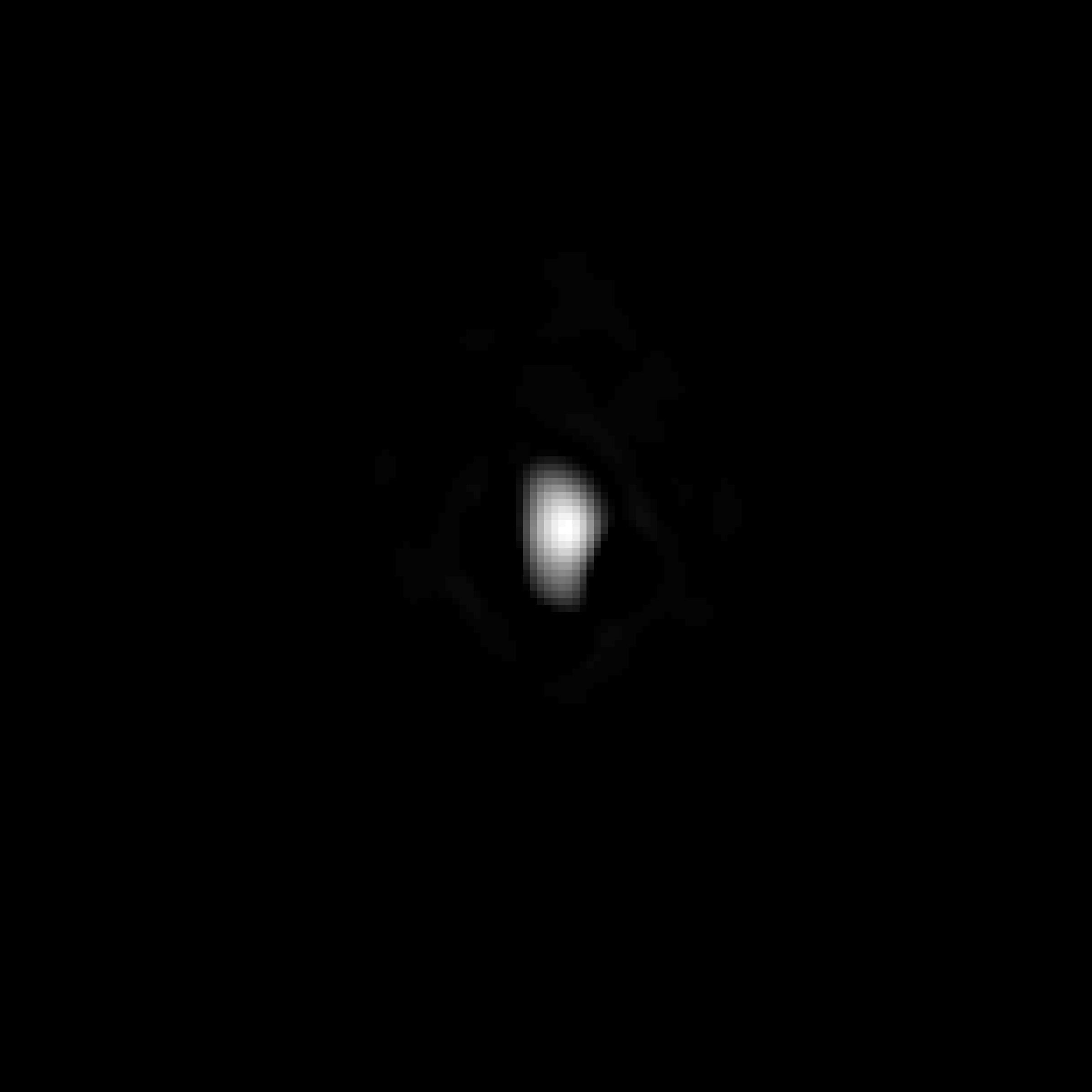}
    \end{subfigure}%
    \begin{subfigure}[b]{0.247\columnwidth}
      \includegraphics[clip=true,trim=90 100 90 100,scale=0.5]{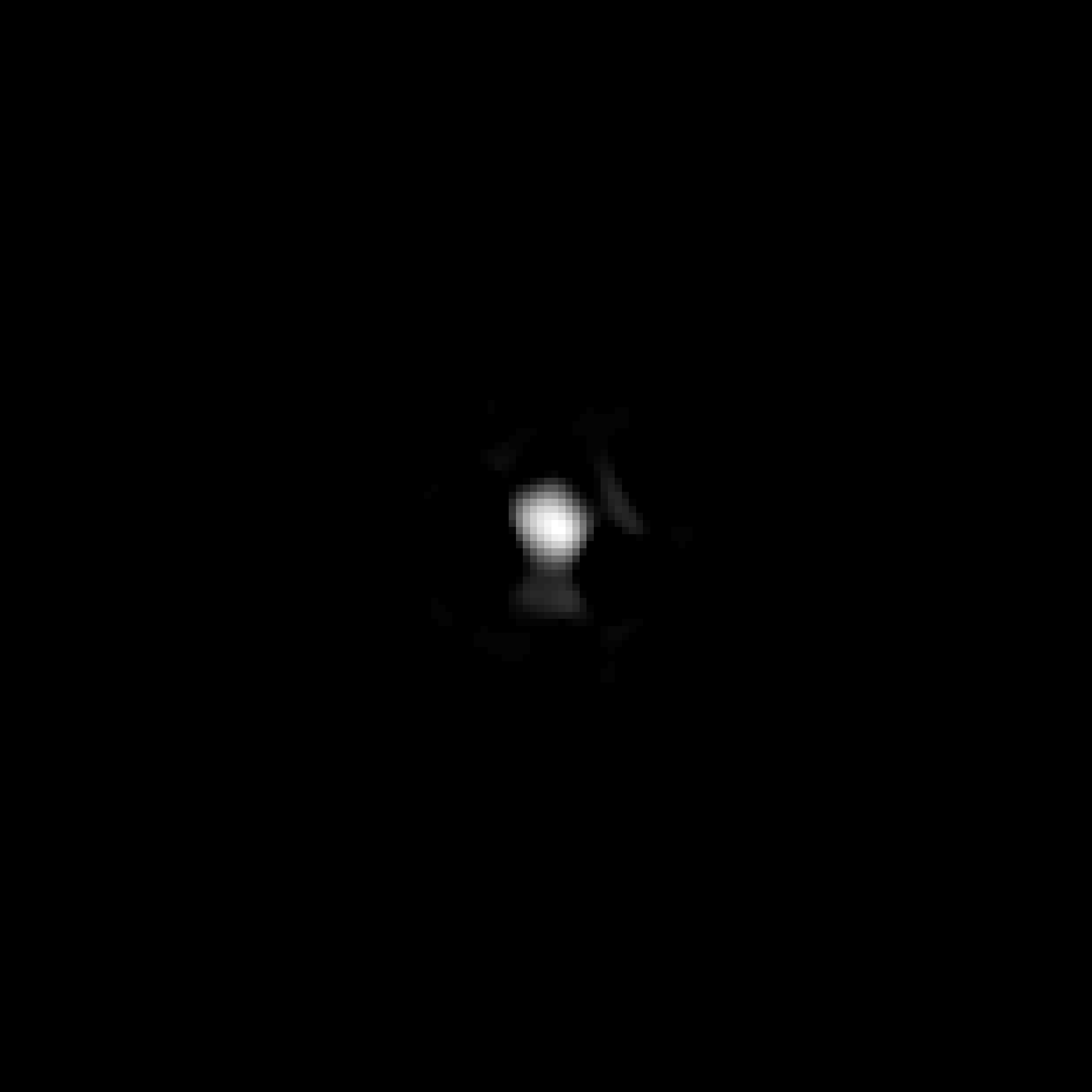}
    \end{subfigure}%
    \begin{subfigure}[b]{0.247\columnwidth}
      \includegraphics[clip=true,trim=90 100 90 100,scale=0.5]{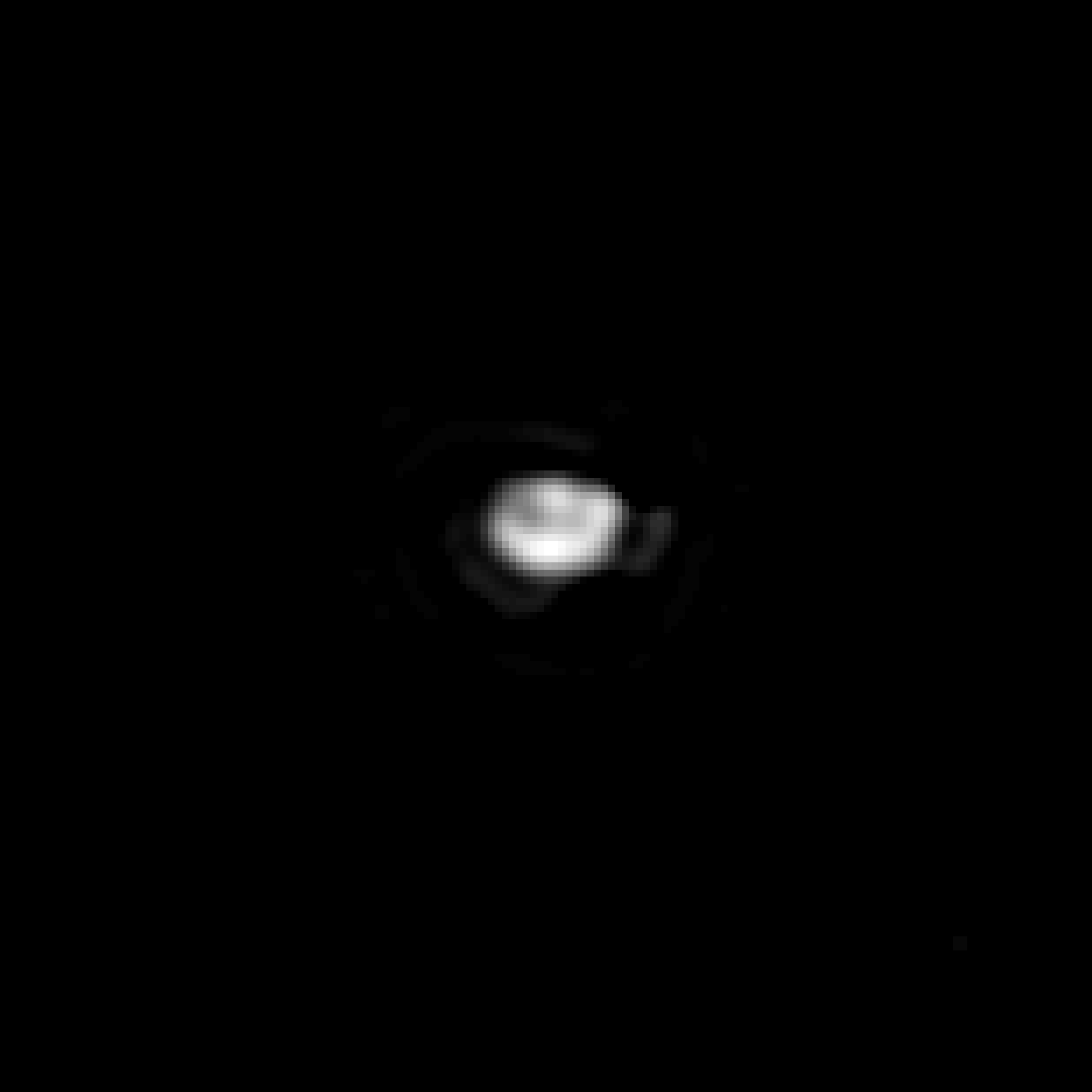}
    \end{subfigure}%
    \begin{subfigure}[b]{0.247\columnwidth}
      \includegraphics[clip=true,trim=90 100 90 100,scale=0.5]{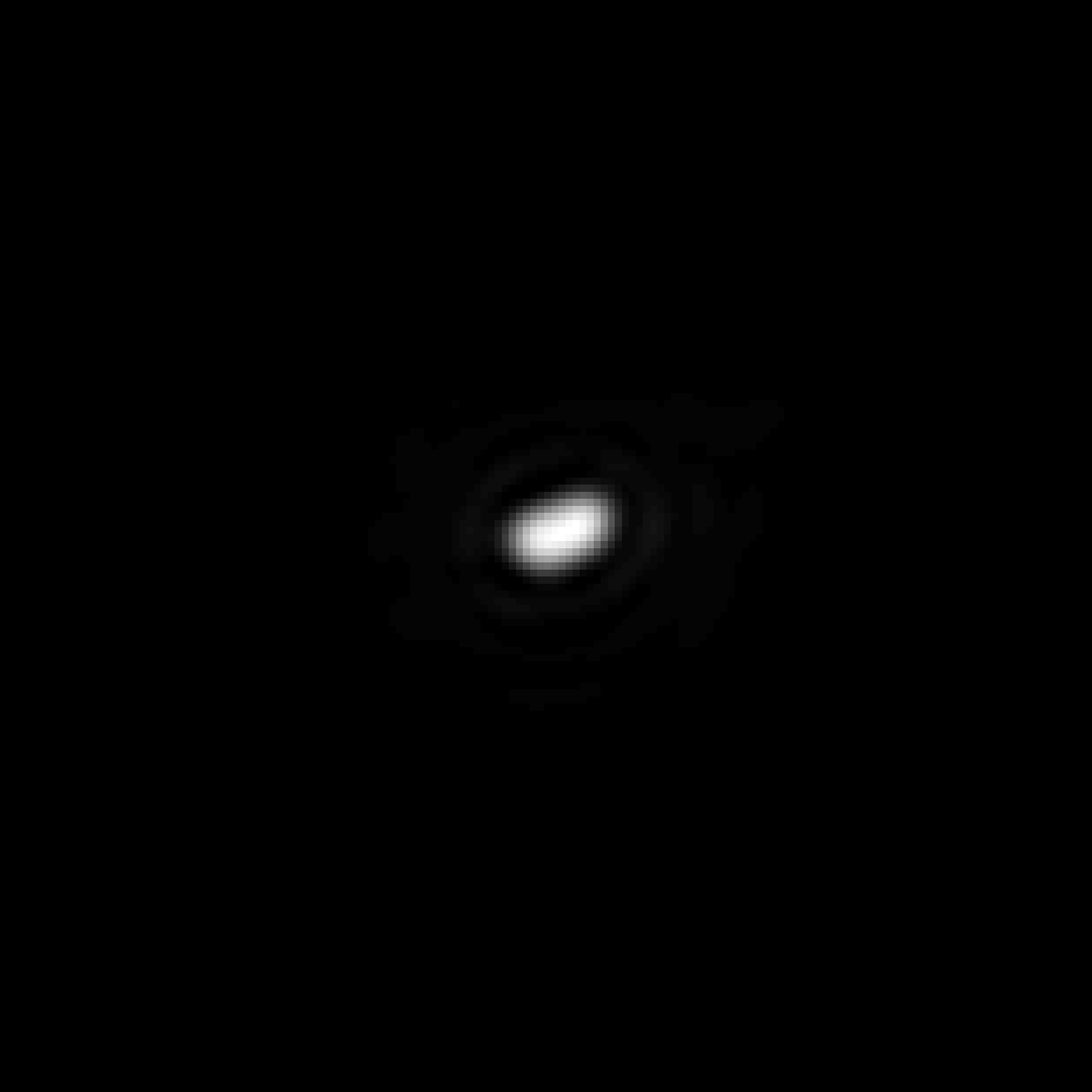}
    \end{subfigure}%
    \begin{subfigure}[b]{0.247\columnwidth}
      \includegraphics[clip=true,trim=90 100 90 100,scale=0.5]{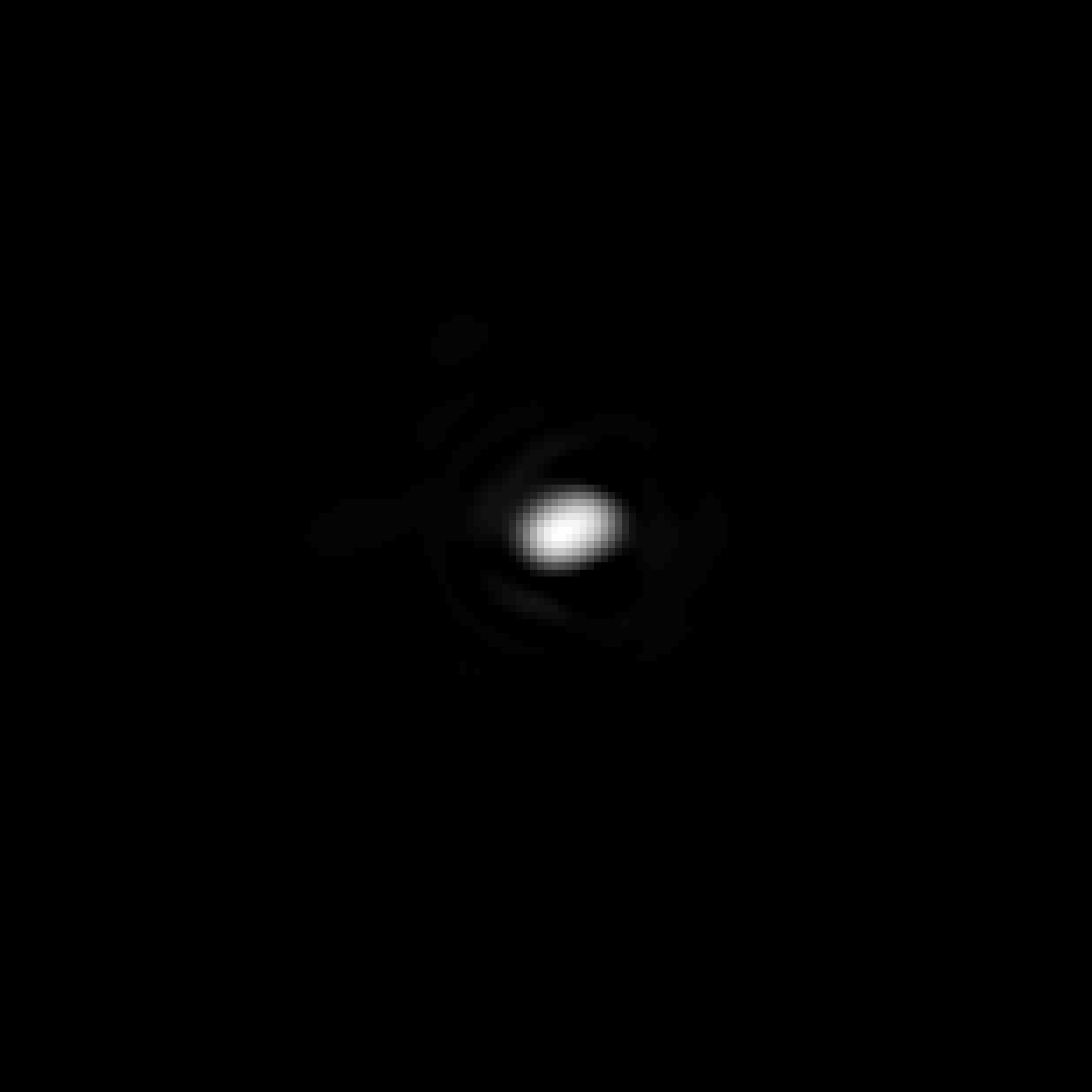}
    \end{subfigure}%
    \begin{subfigure}[b]{0.247\columnwidth}
      \includegraphics[clip=true,trim=90 100 90 100,scale=0.5]{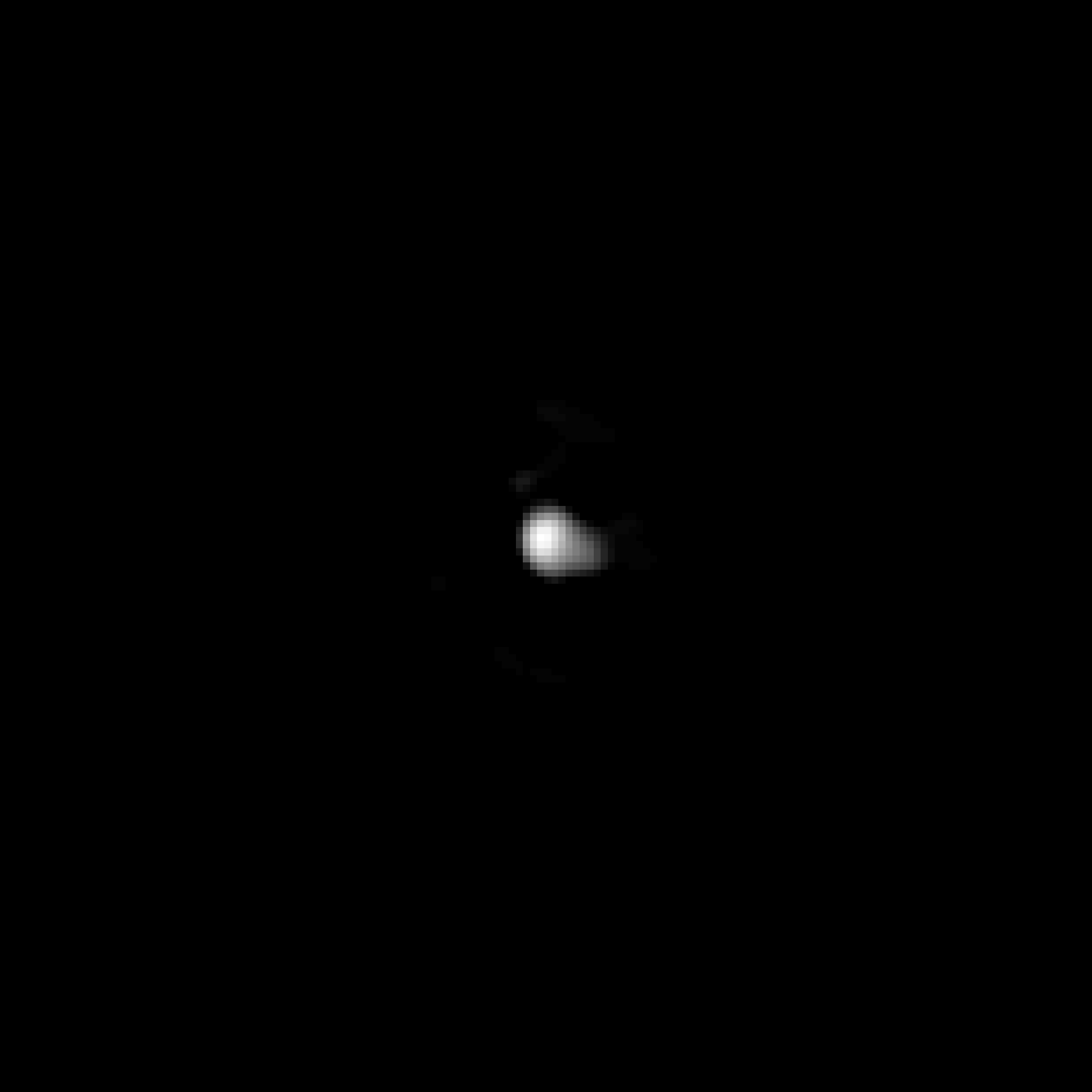}
    \end{subfigure}%

    \begin{subfigure}[b]{0.247\columnwidth}
      \includegraphics[clip=true,trim=90 100 90 100,scale=0.5]{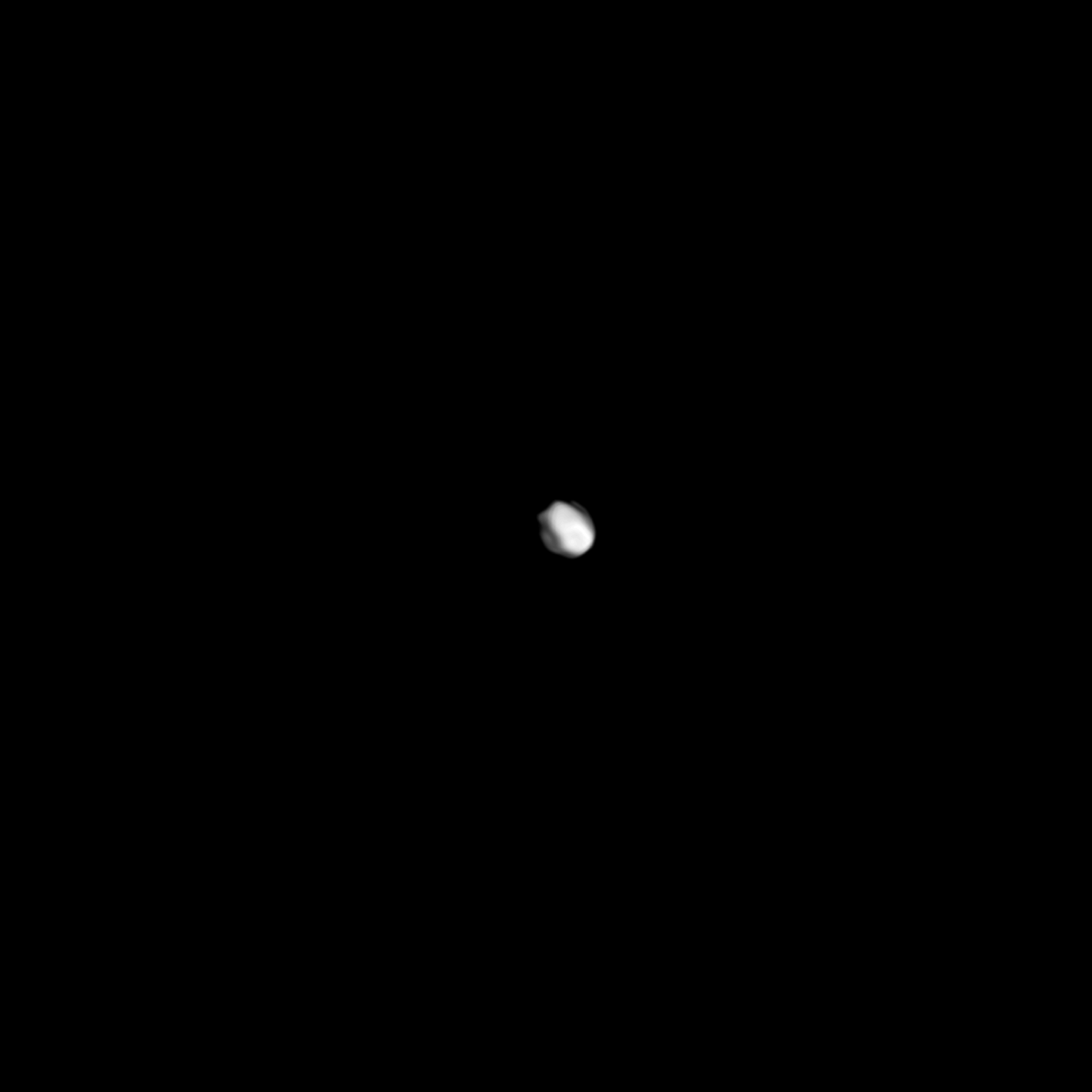}
    \end{subfigure}%
    \begin{subfigure}[b]{0.247\columnwidth}
      \includegraphics[clip=true,trim=90 100 90 100,scale=0.5]{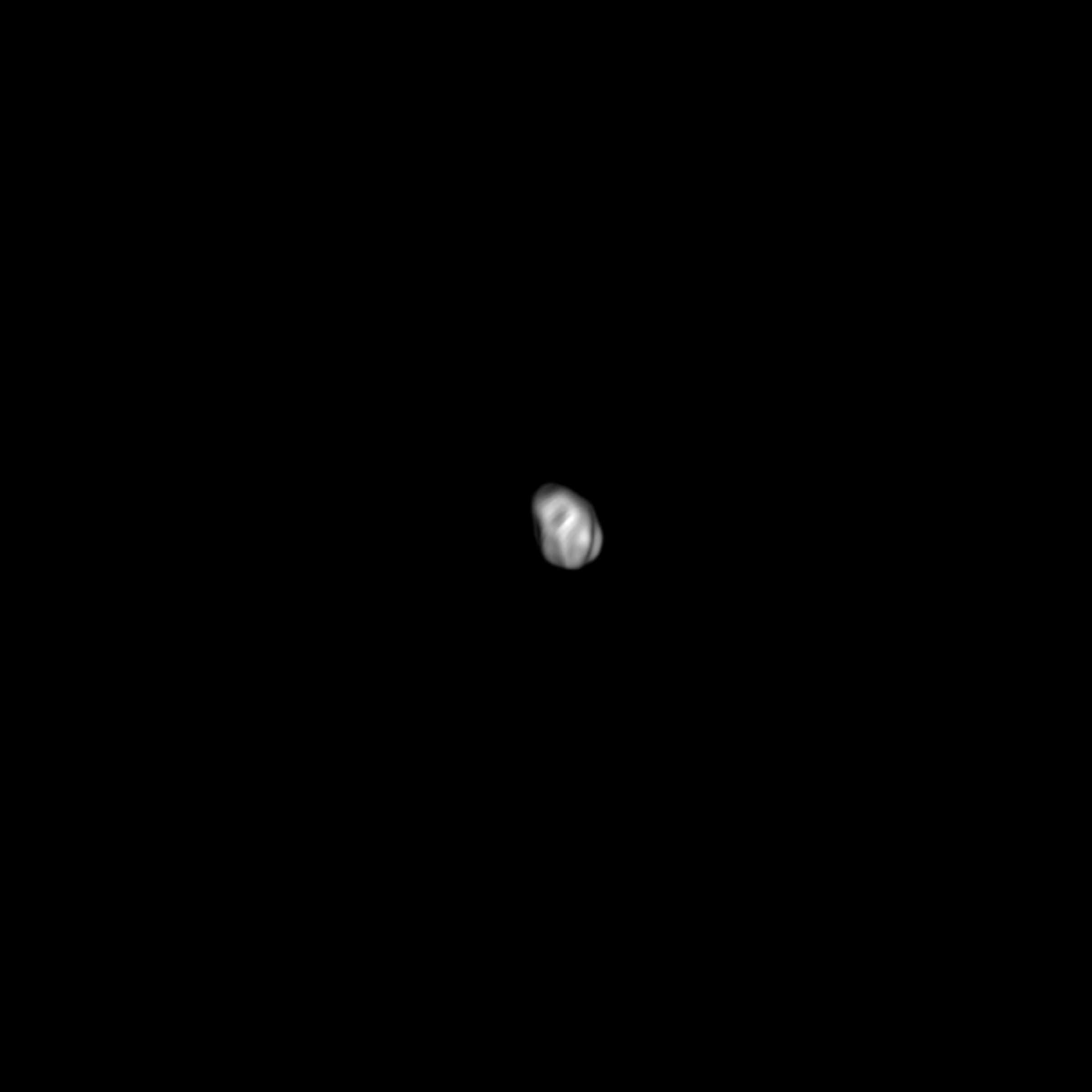}
    \end{subfigure}%
    \begin{subfigure}[b]{0.247\columnwidth}
      \includegraphics[clip=true,trim=90 100 90 100,scale=0.5]{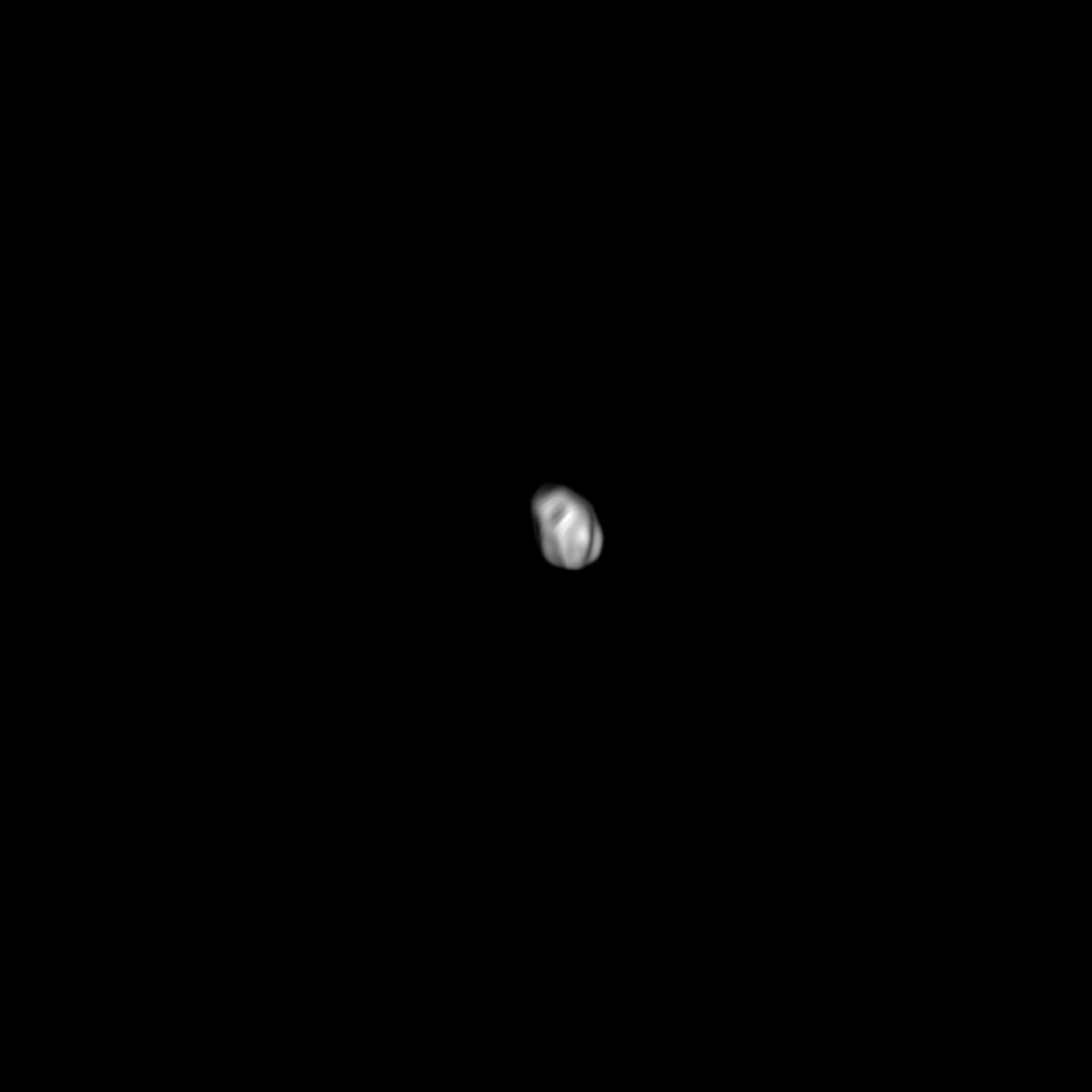}
    \end{subfigure}%
    \begin{subfigure}[b]{0.247\columnwidth}
      \includegraphics[clip=true,trim=90 100 90 100,scale=0.5]{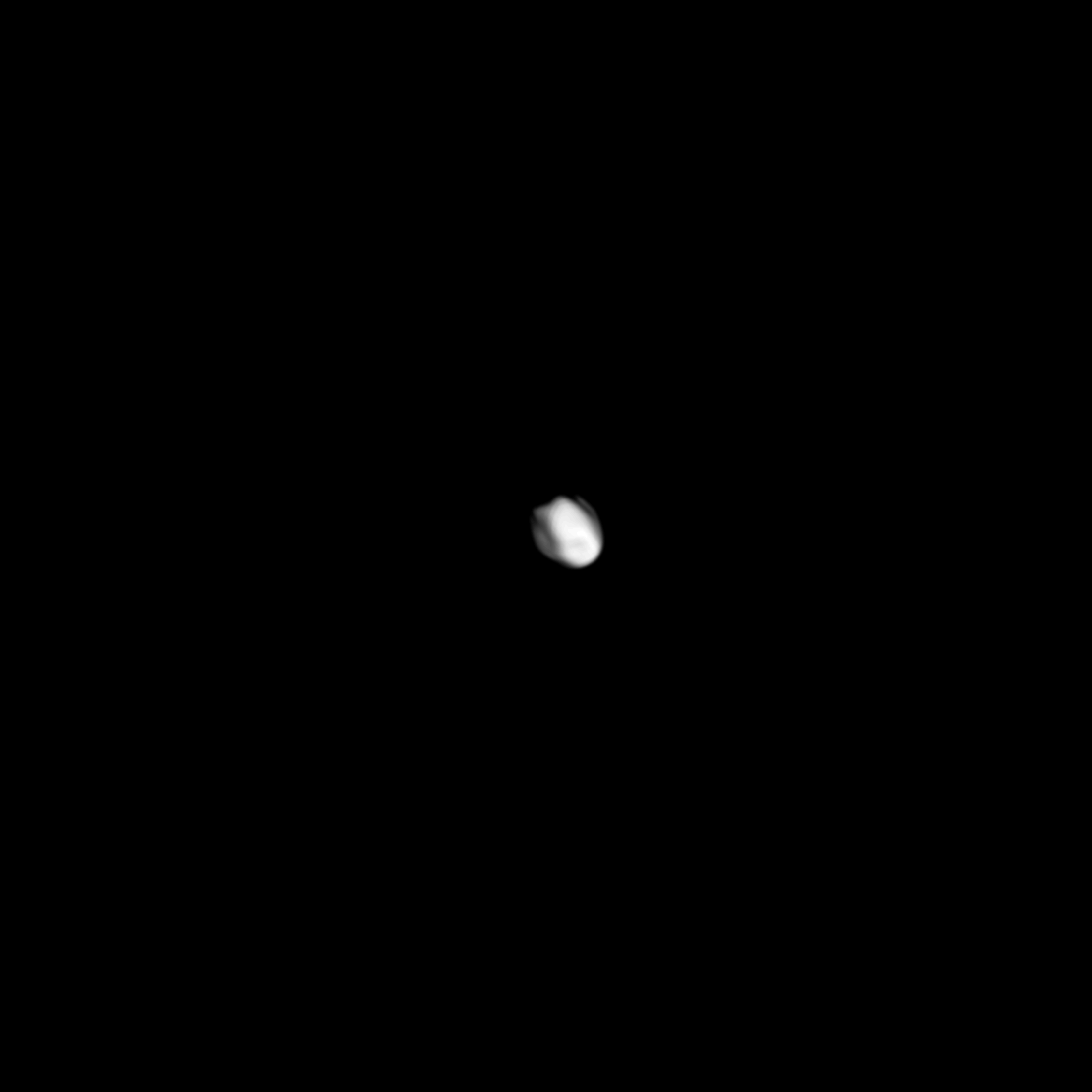}
    \end{subfigure}%
    \begin{subfigure}[b]{0.247\columnwidth}
      \includegraphics[clip=true,trim=90 100 90 100,scale=0.5]{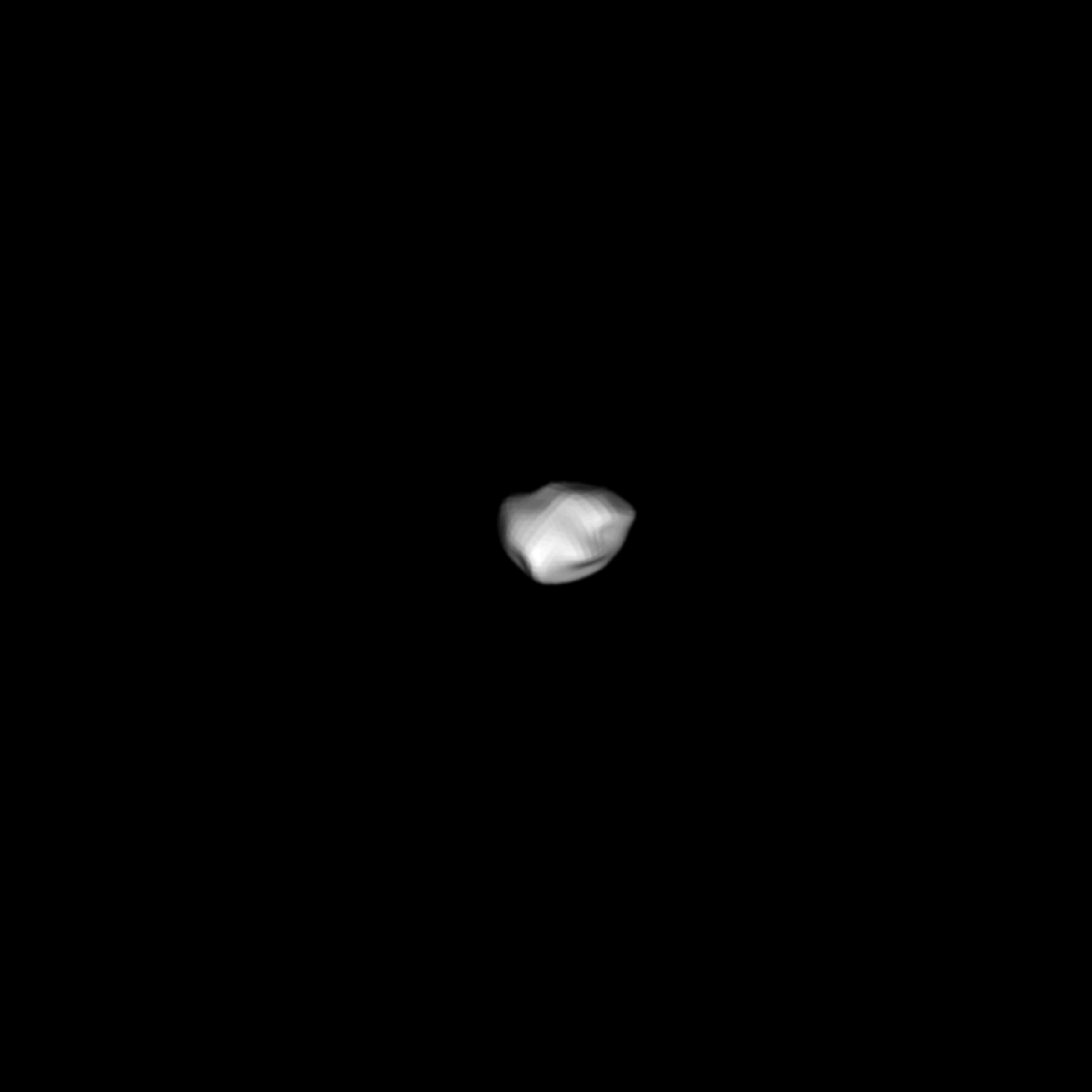}
    \end{subfigure}%
    \begin{subfigure}[b]{0.247\columnwidth}
      \includegraphics[clip=true,trim=90 100 90 100,scale=0.5]{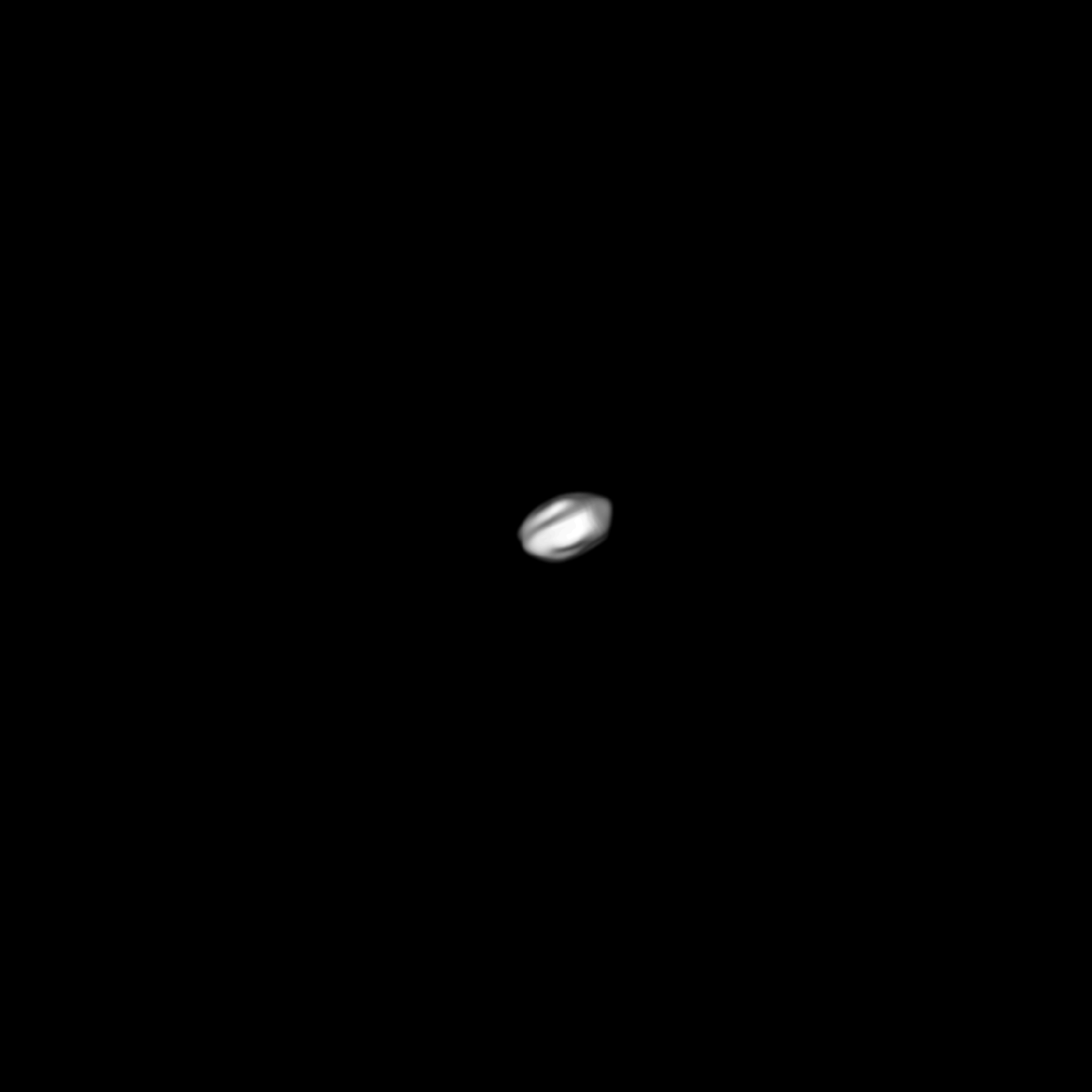}
    \end{subfigure}%
    \begin{subfigure}[b]{0.247\columnwidth}
      \includegraphics[clip=true,trim=90 100 90 100,scale=0.5]{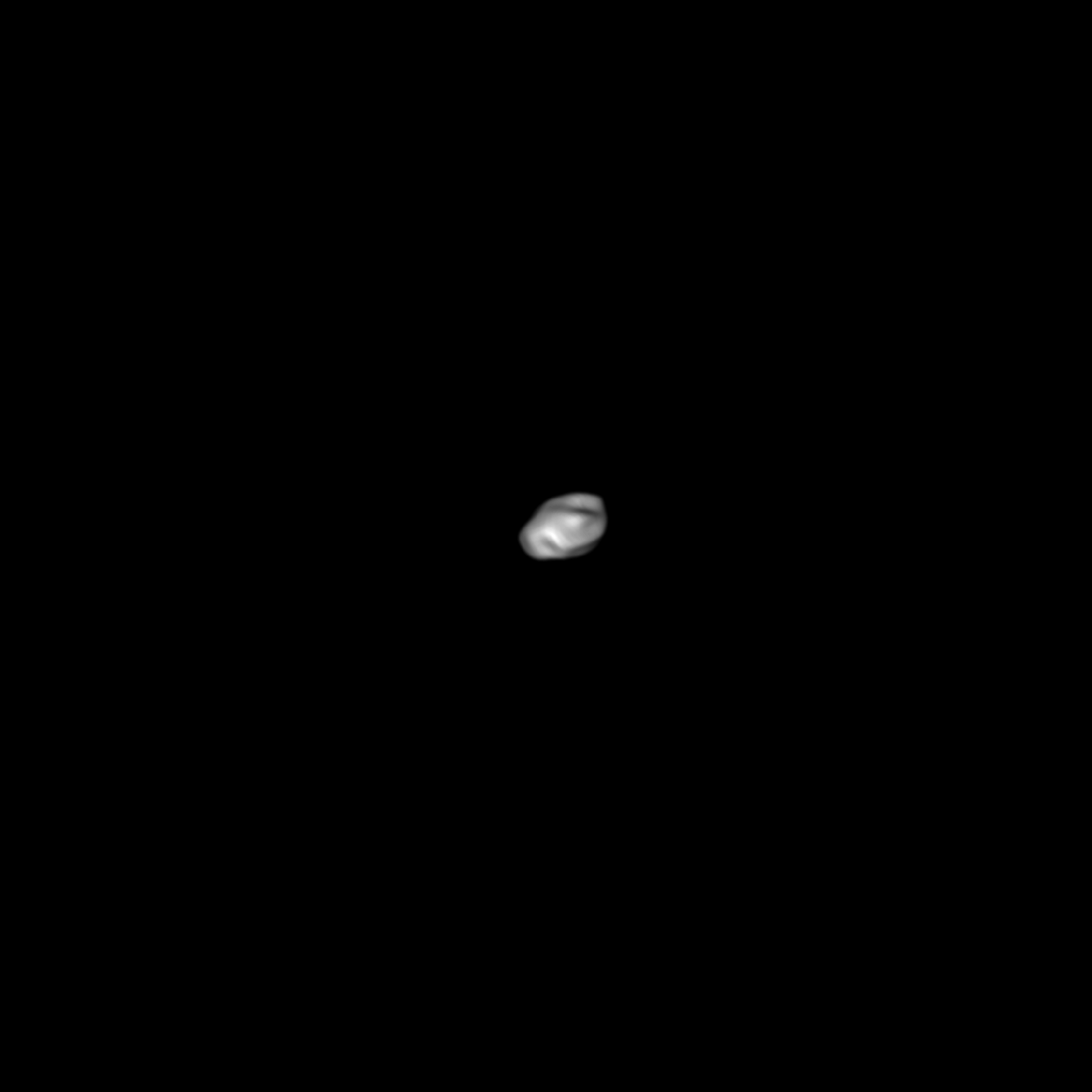}
    \end{subfigure}%
    \begin{subfigure}[b]{0.247\columnwidth}
      \includegraphics[clip=true,trim=90 100 90 100,scale=0.5]{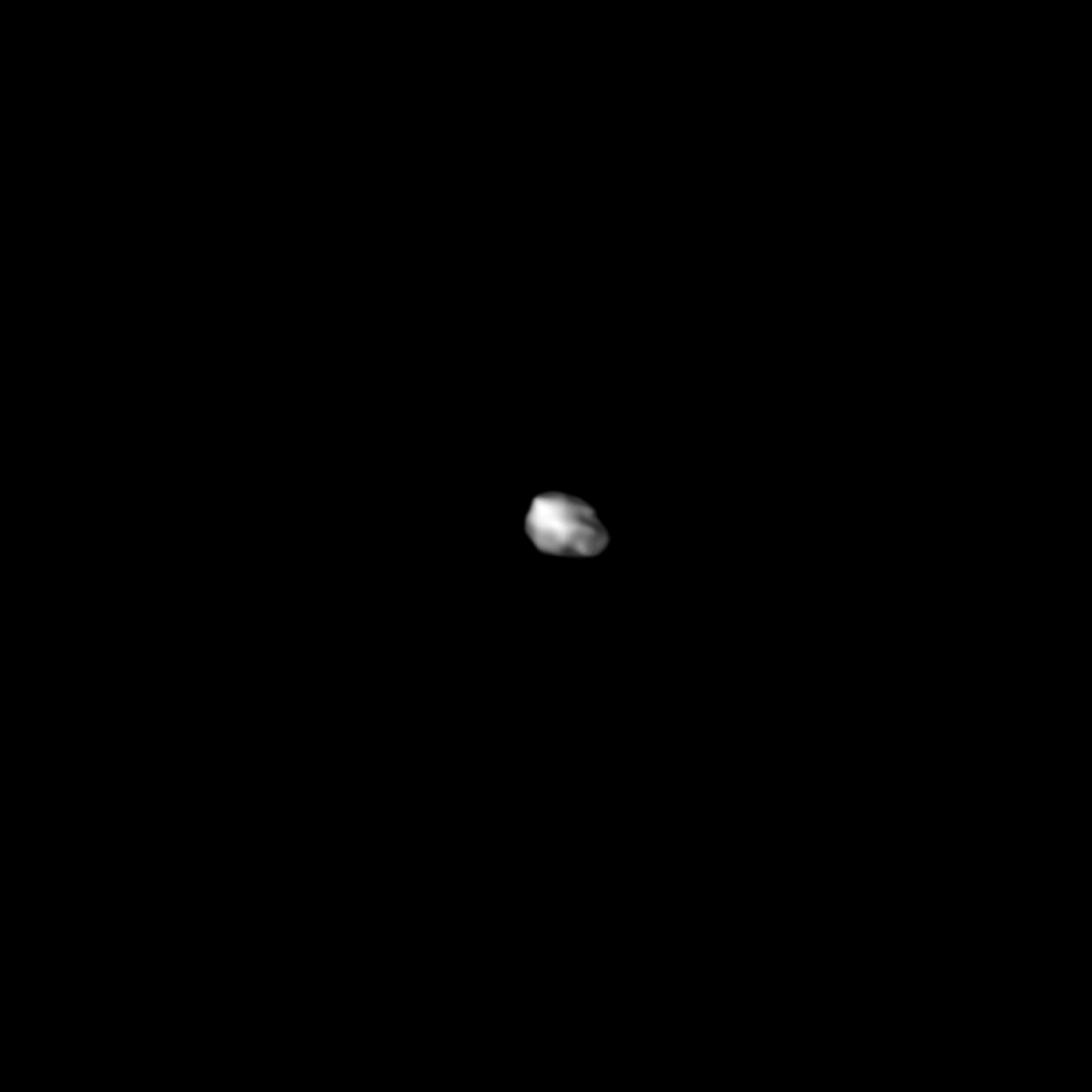}
    \end{subfigure}%
    
    \begin{subfigure}[b]{0.247\columnwidth}
      \includegraphics[clip=true,trim=90 100 90 100,scale=0.5]{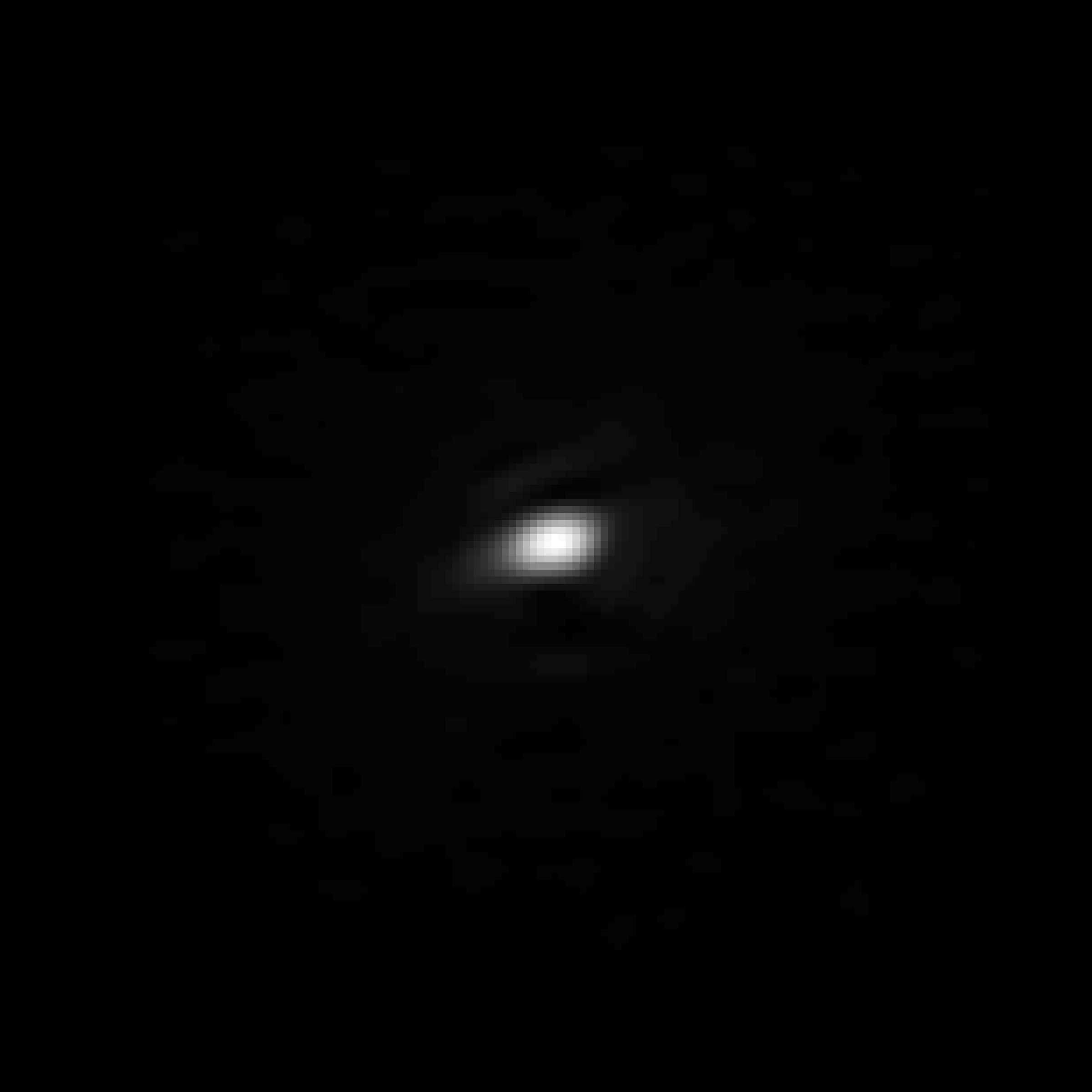}
    \end{subfigure}%
    \begin{subfigure}[b]{0.247\columnwidth}
      \includegraphics[clip=true,trim=90 100 90 100,scale=0.5]{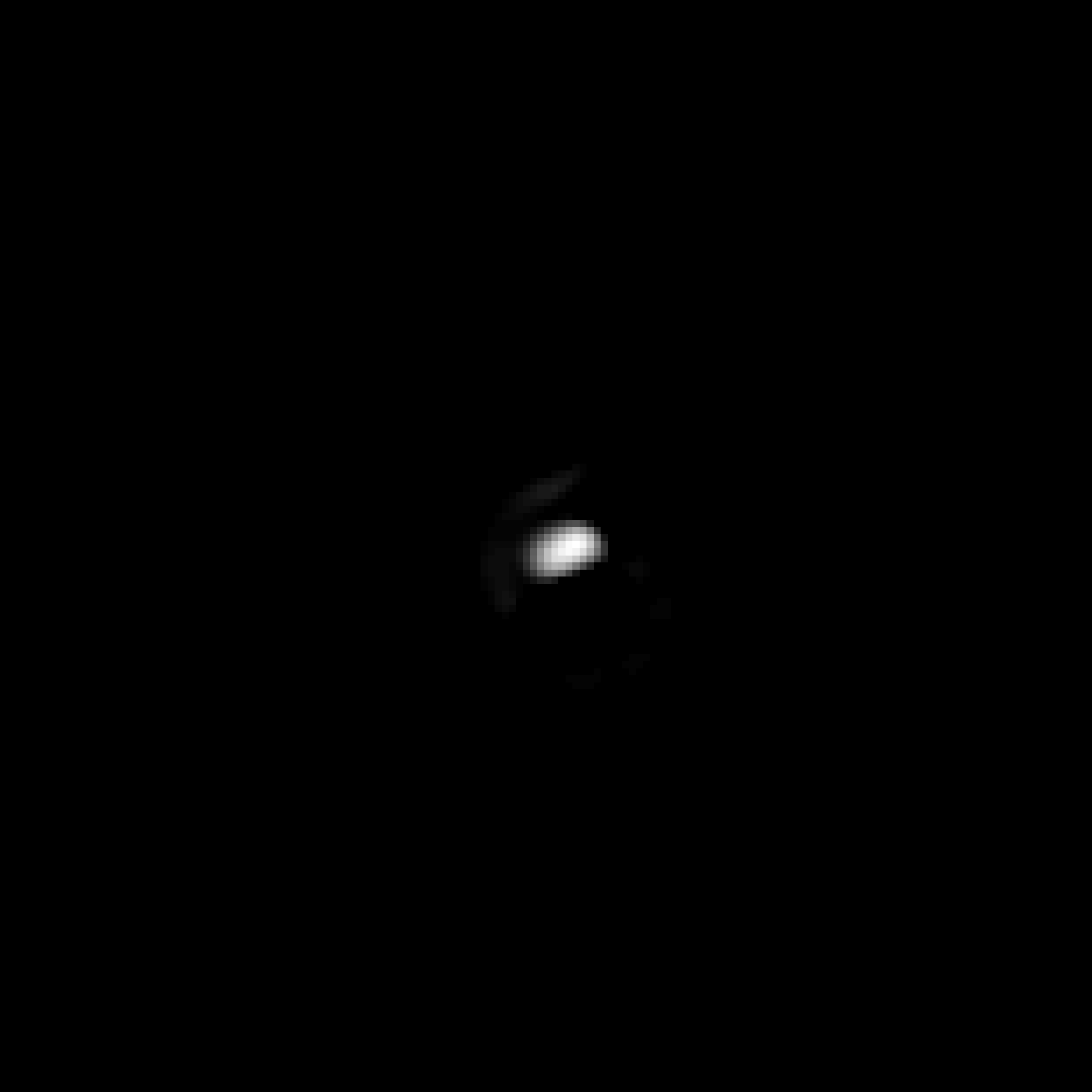}
    \end{subfigure}%
    \begin{subfigure}[b]{0.247\columnwidth}
      \includegraphics[clip=true,trim=90 100 90 100,scale=0.5]{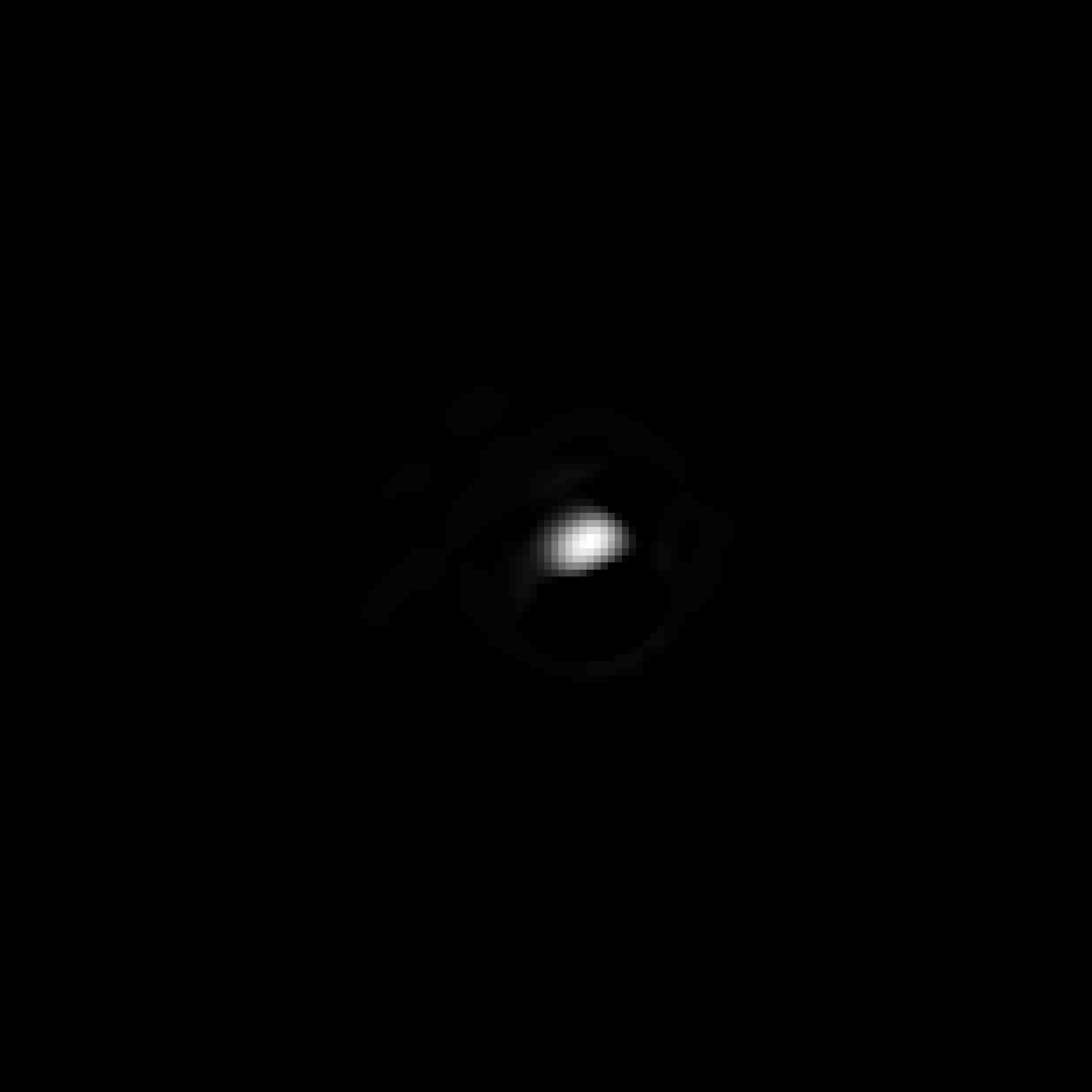}
    \end{subfigure}%
    \begin{subfigure}[b]{0.247\columnwidth}
      \includegraphics[clip=true,trim=90 100 90 100,scale=0.5]{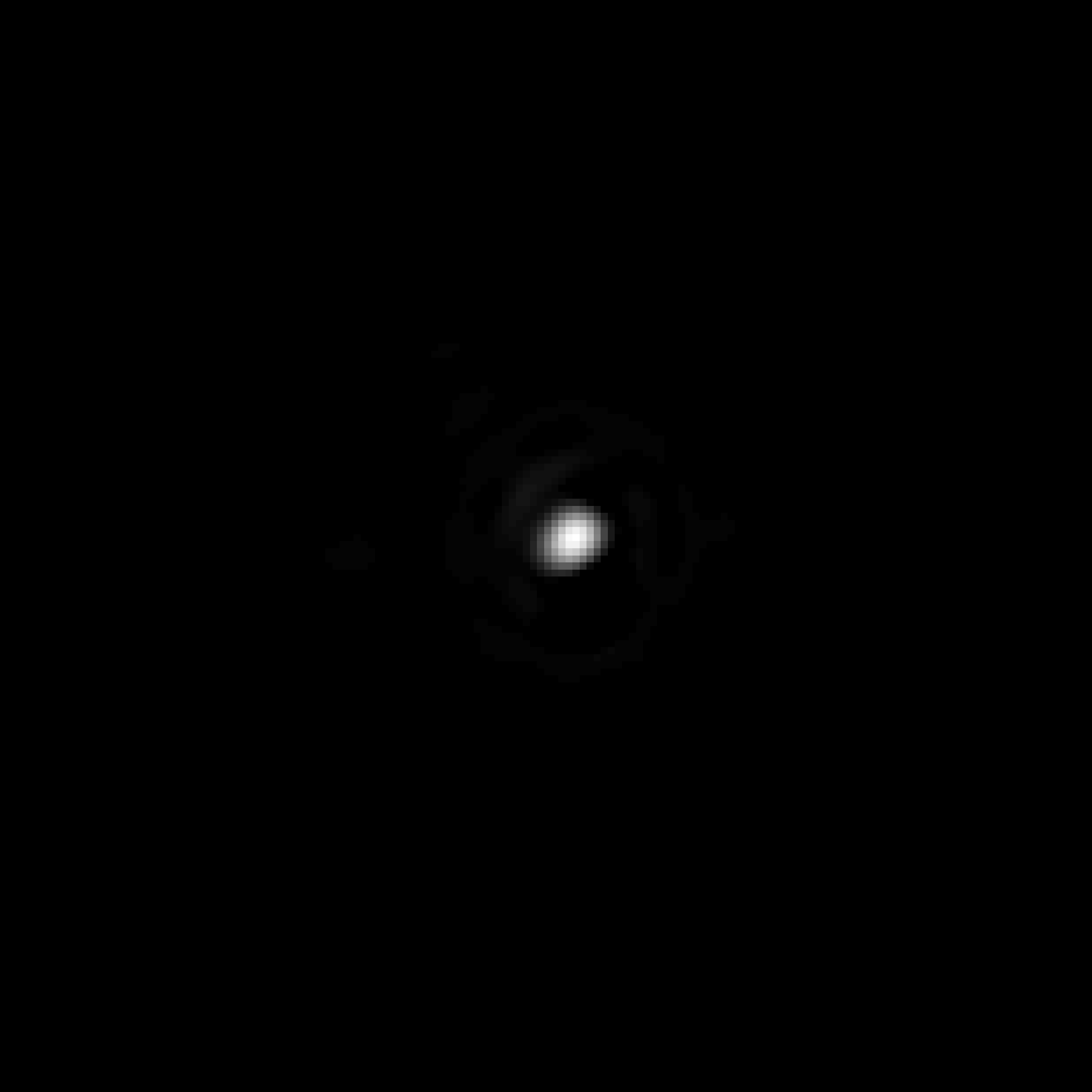}
    \end{subfigure}%
    \begin{subfigure}[b]{0.247\columnwidth}
      \includegraphics[clip=true,trim=90 100 90 100,scale=0.5]{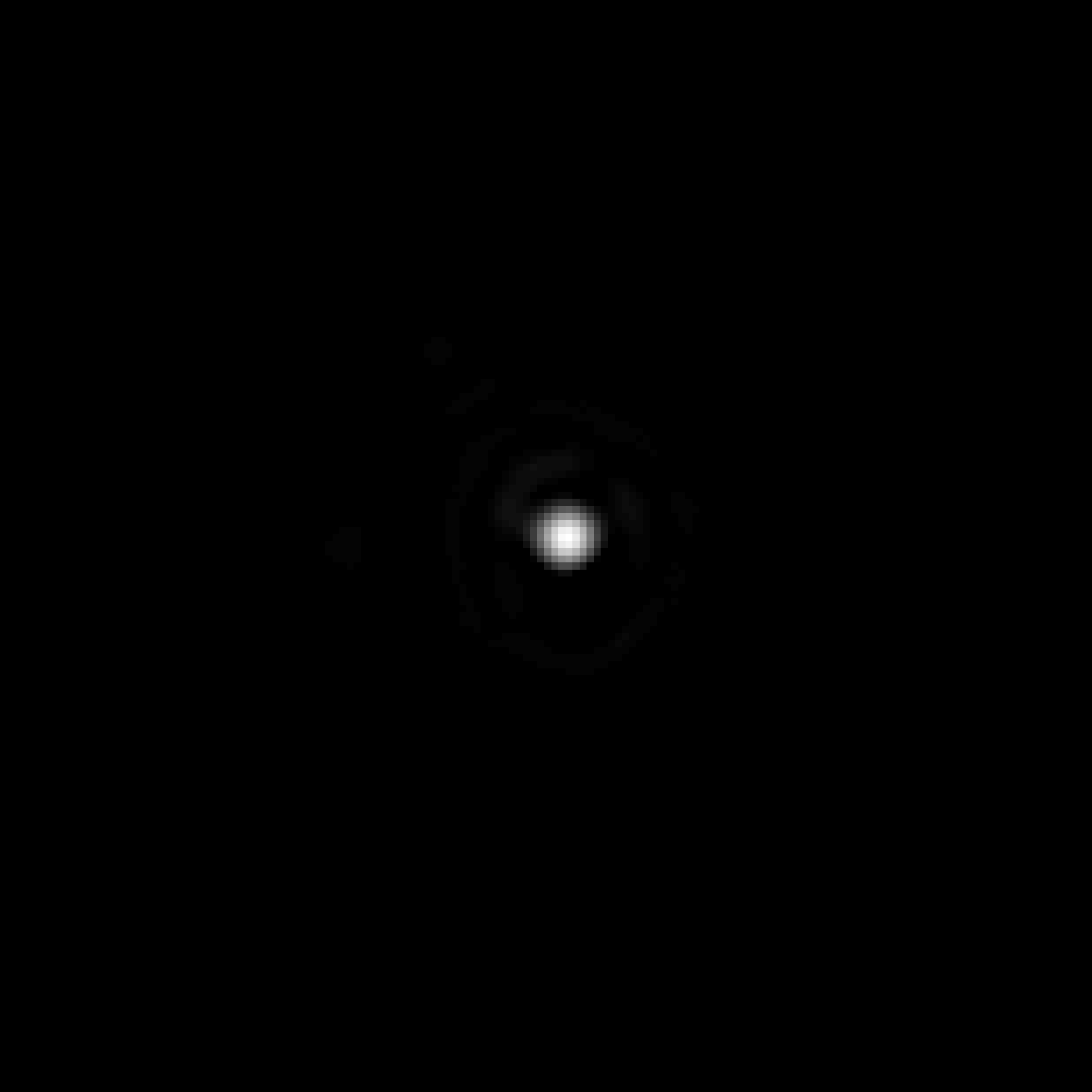}
    \end{subfigure}%
    \begin{subfigure}[b]{0.247\columnwidth}
      \includegraphics[clip=true,trim=90 100 90 100,scale=0.5]{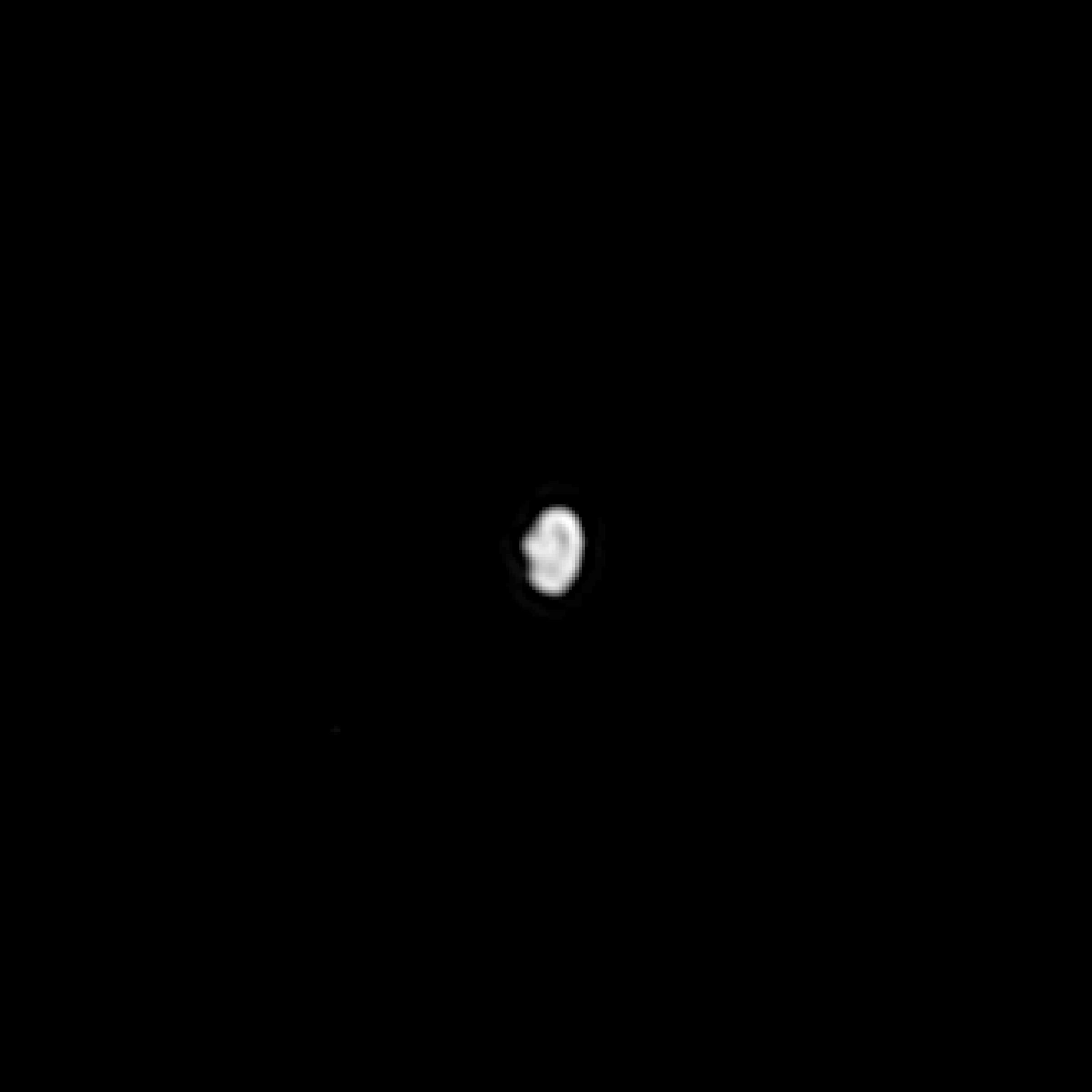}
    \end{subfigure}%
    \begin{subfigure}[b]{0.247\columnwidth}
      \includegraphics[clip=true,trim=90 100 90 100,scale=0.5]{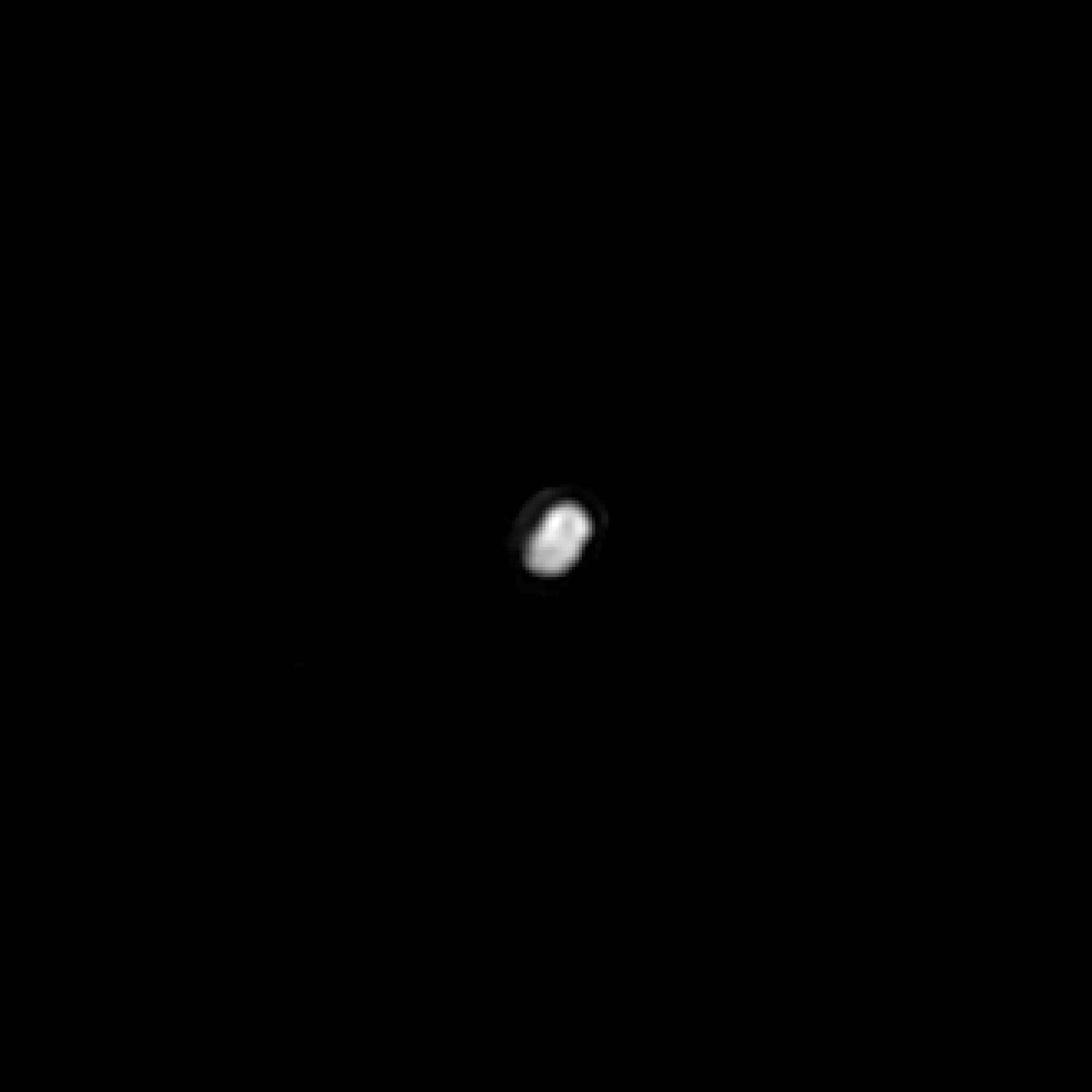}
    \end{subfigure}%

    \begin{subfigure}[b]{0.247\columnwidth}
      \includegraphics[clip=true,trim=90 100 90 100,scale=0.5]{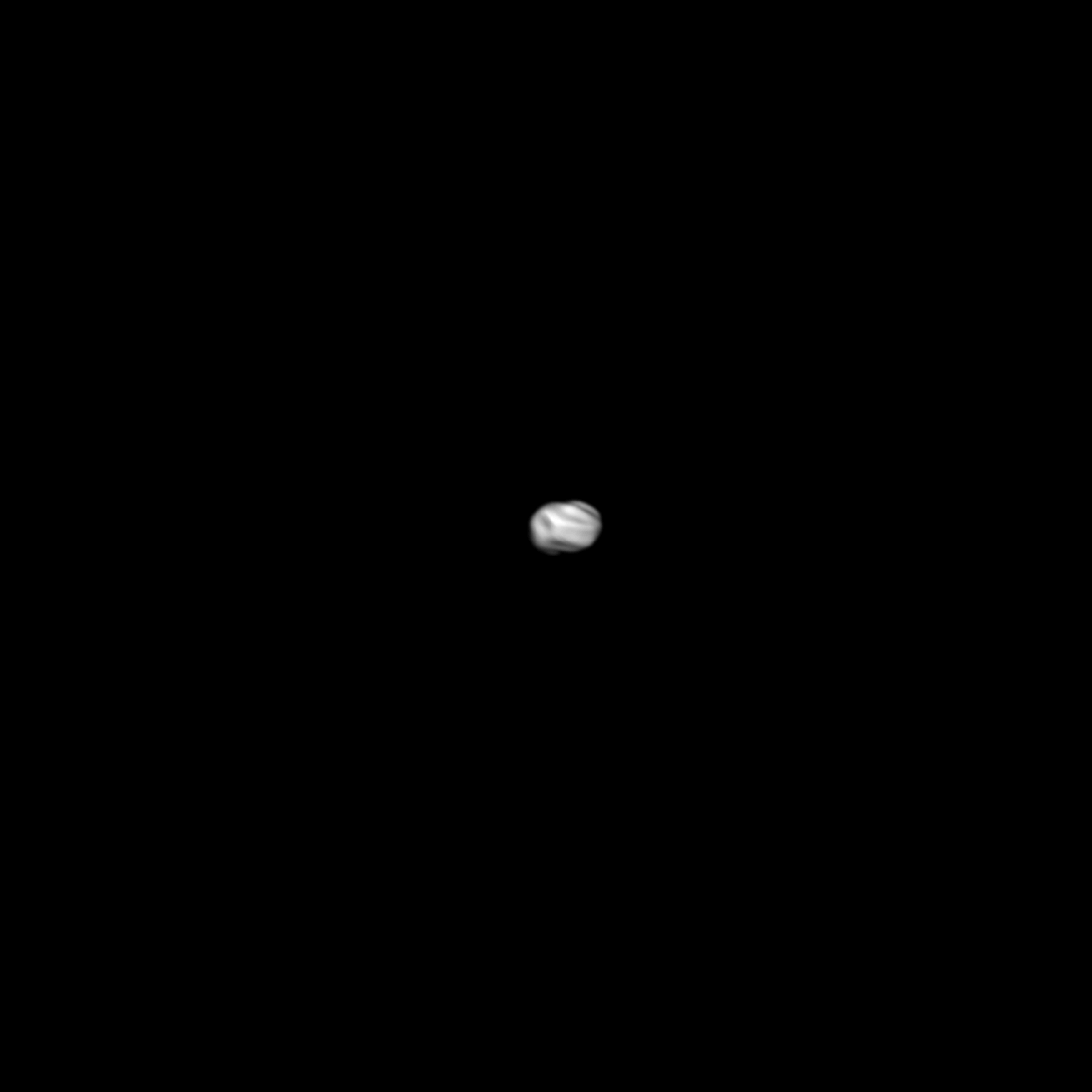}
    \end{subfigure}%
    \begin{subfigure}[b]{0.247\columnwidth}
      \includegraphics[clip=true,trim=90 100 90 100,scale=0.5]{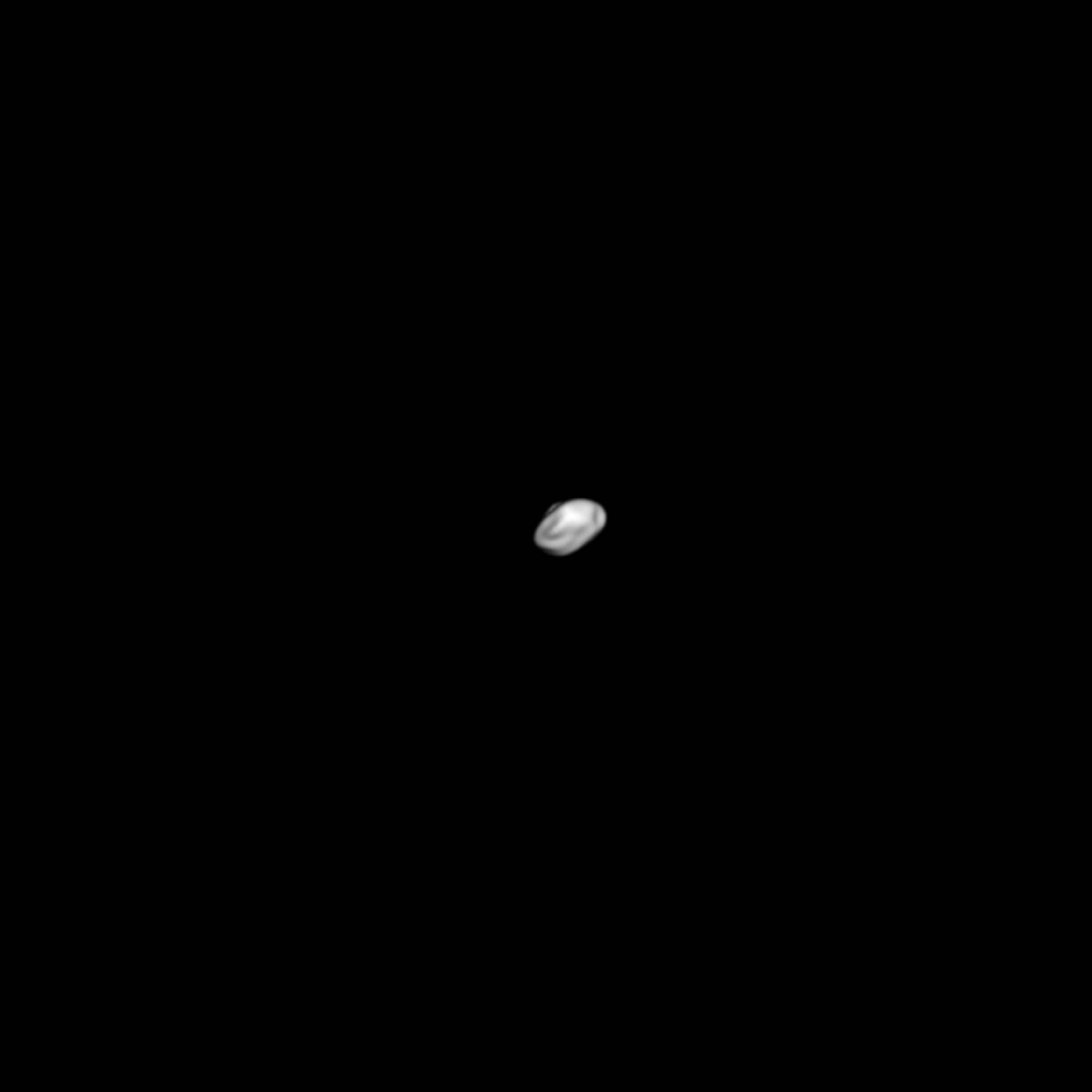}
    \end{subfigure}%
    \begin{subfigure}[b]{0.247\columnwidth}
      \includegraphics[clip=true,trim=90 100 90 100,scale=0.5]{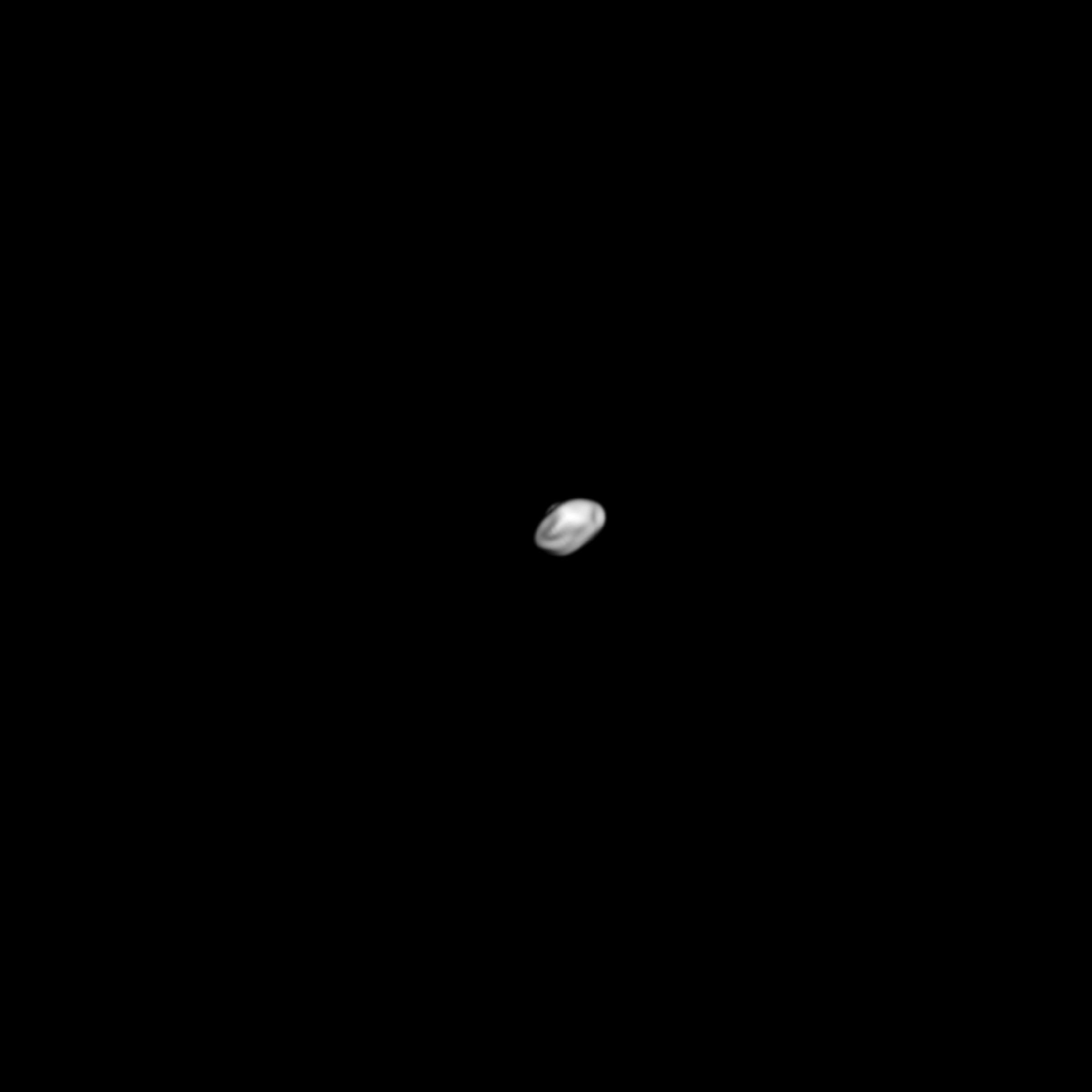}
    \end{subfigure}%
    \begin{subfigure}[b]{0.247\columnwidth}
      \includegraphics[clip=true,trim=90 100 90 100,scale=0.5]{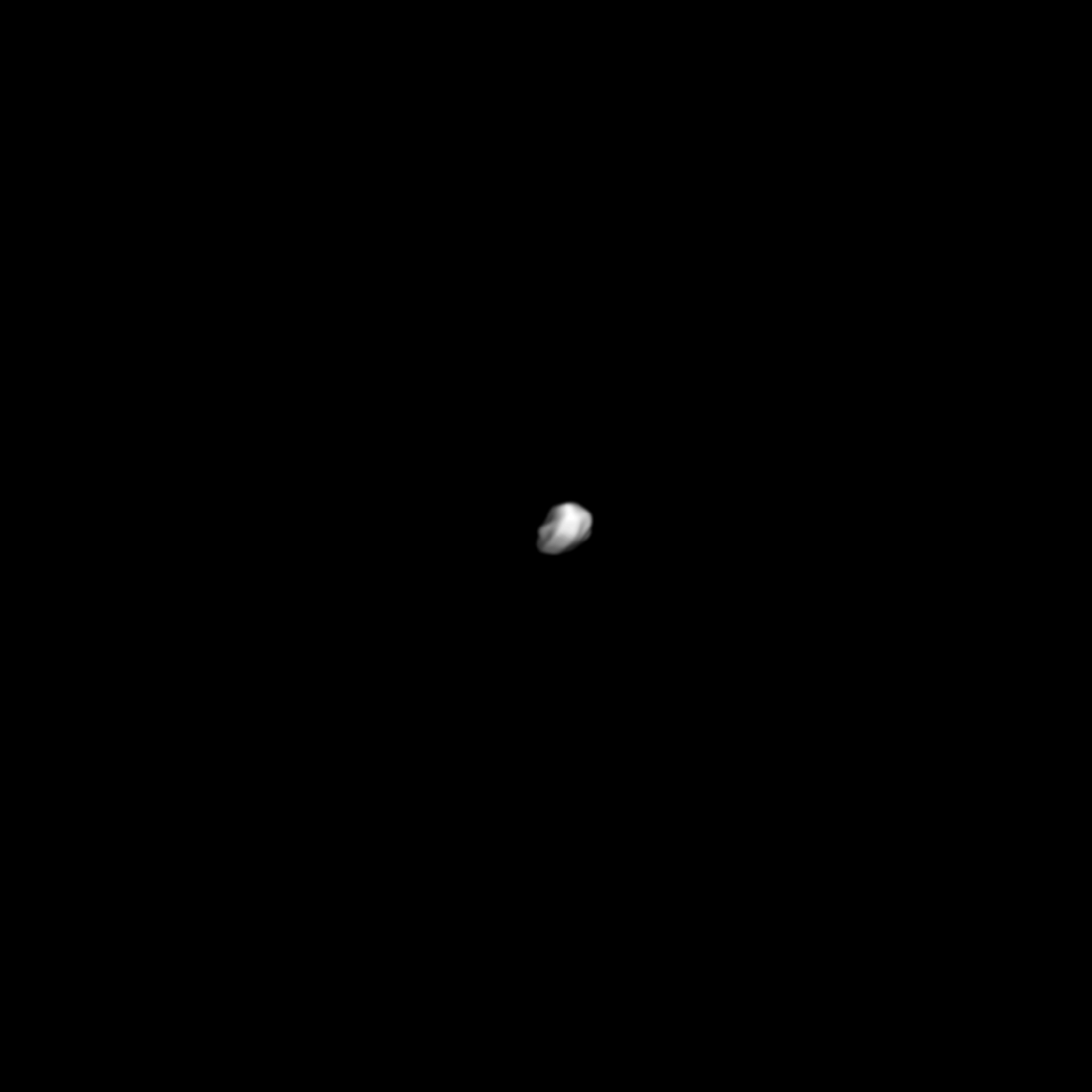}
    \end{subfigure}%
    \begin{subfigure}[b]{0.247\columnwidth}
      \includegraphics[clip=true,trim=90 100 90 100,scale=0.5]{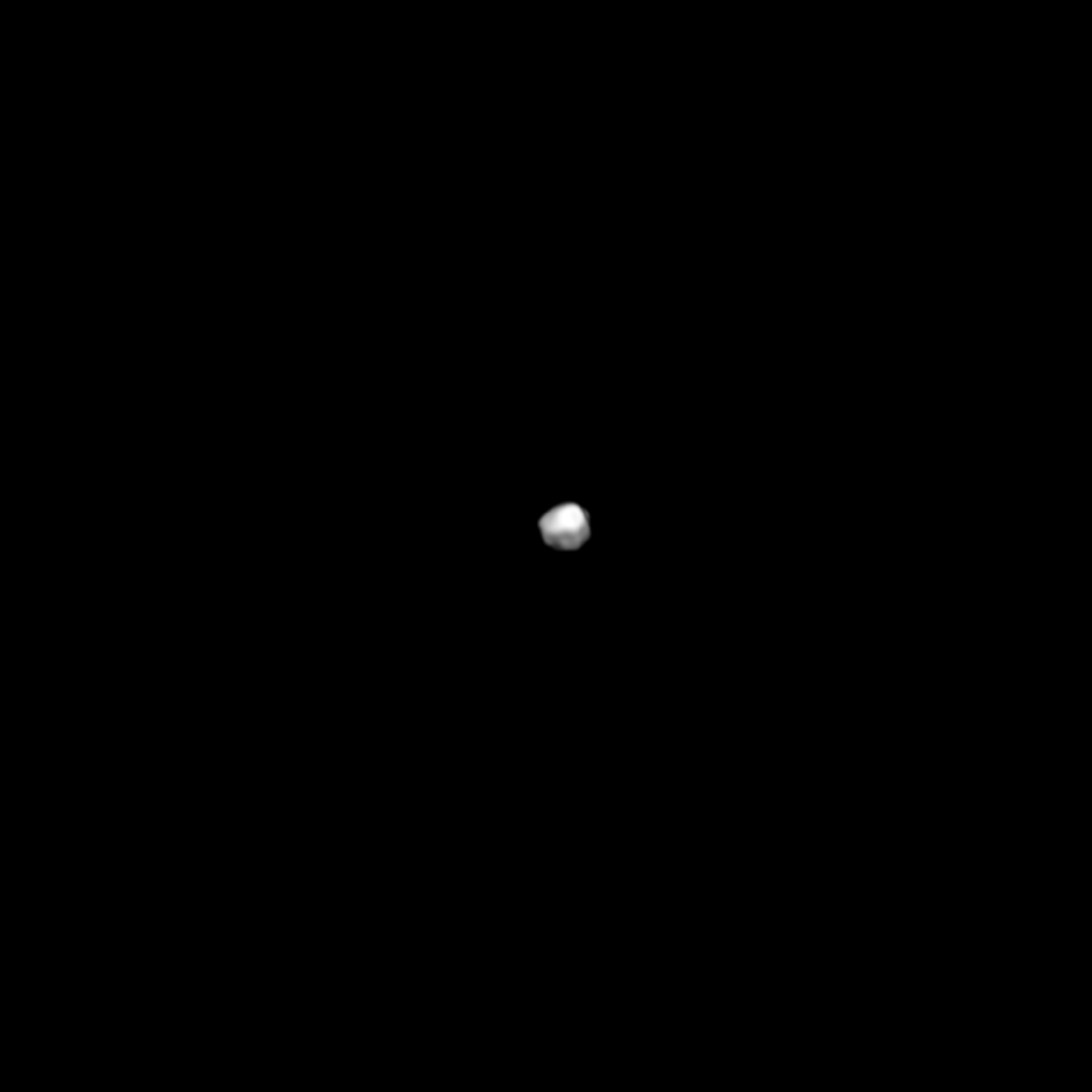}
    \end{subfigure}%
    \begin{subfigure}[b]{0.247\columnwidth}
      \includegraphics[clip=true,trim=90 100 90 100,scale=0.5]{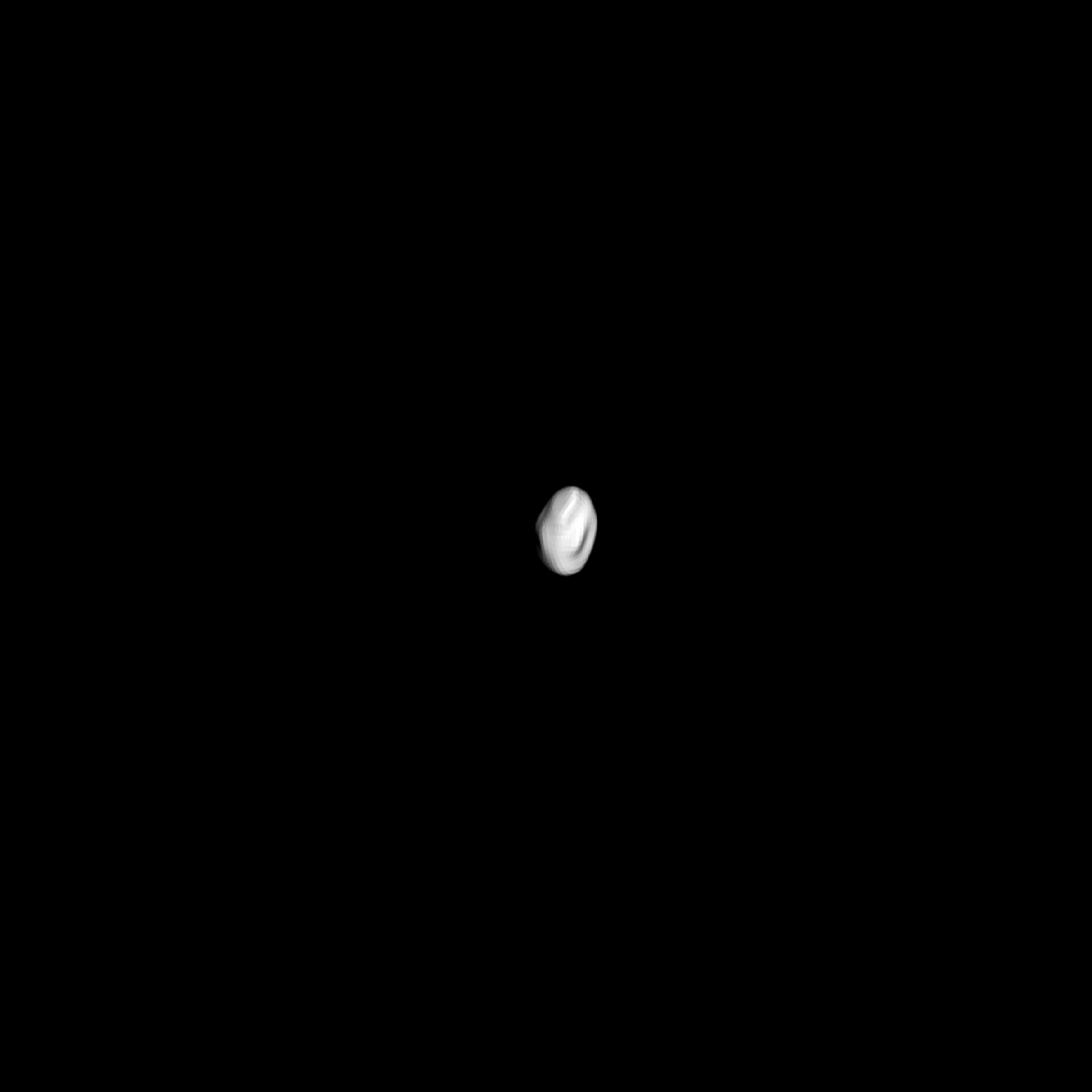}
    \end{subfigure}%
    \begin{subfigure}[b]{0.247\columnwidth}
      \includegraphics[clip=true,trim=90 100 90 100,scale=0.5]{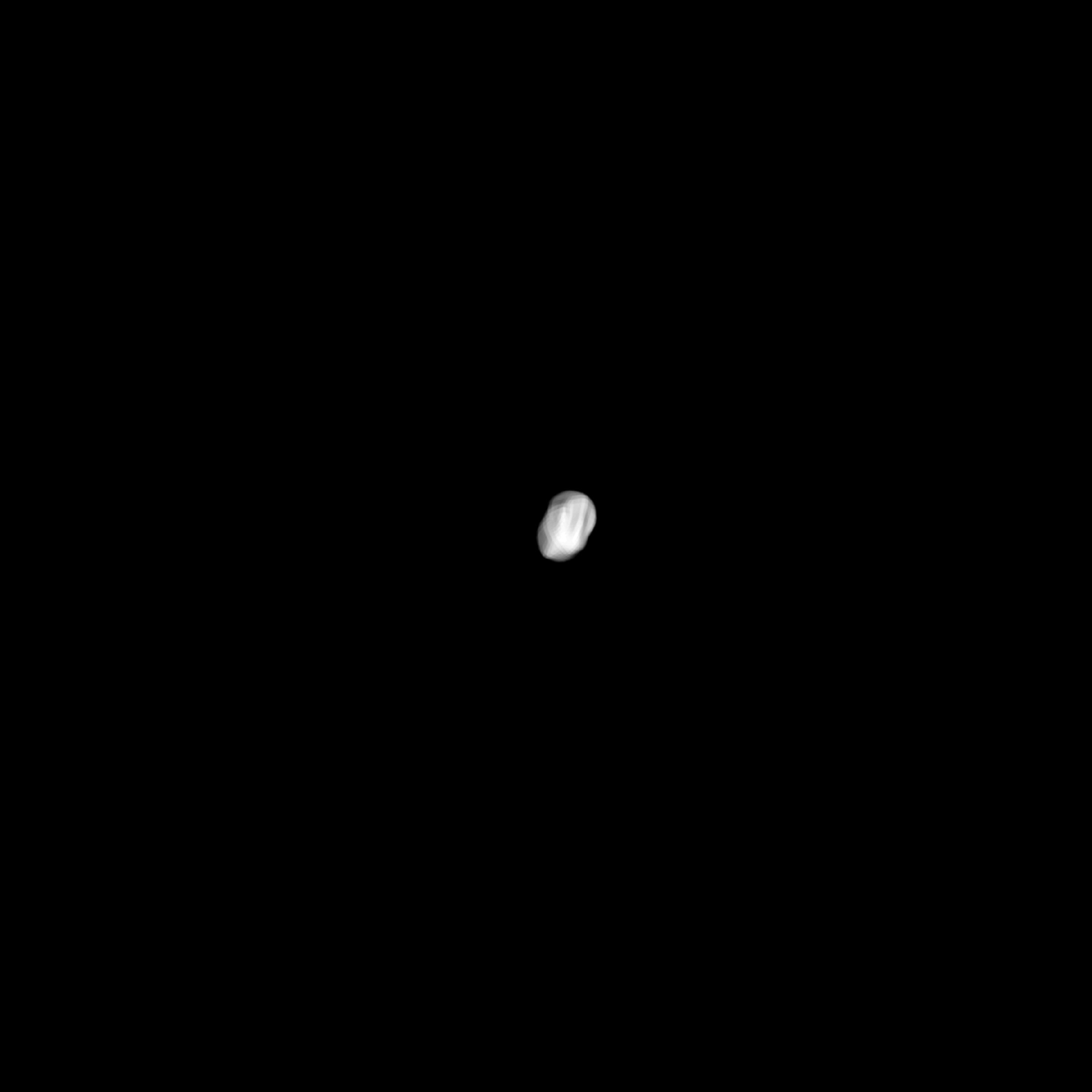}
    \end{subfigure}%
     \caption{\label{fig:comparison}Deconvolved images of Elektra (top) and corresponding model (bottom). First thirteen images are obtained by the Keck Nirc2 and the last two by the VLT/SPHERE instrument.}
\end{figure*}

ADAM allows the usage of two different shape supports -- subdivision surfaces and octanoids. Moreover, we can either use the raw AO images together with the point-spread function, or the deconvolved images alone. This gave us four different combinations for the shape modeling, which were used for shape and size uncertainty assessment. On top of that, we also tested assigning different weights to the AO data with respect to the light curve data, as well as assigning different weights to individual images. For example, better resolved images were weighted more, especially those from SPHERE. This approach further constrained the size and its uncertainty.

First, we modeled Elektra's shape from optical light curves and resolved images from SPHERE, because the SPHERE images contain more detailed information about the shape than the images from Nirc2 due to their higher spacial resolution. This shape model contains multiple features included in the data, although some of them might be artificial. The shape model of Elektra reconstructed from disk-integrated optical data and raw SPHERE images with the subdivision surfaces shape support is shown in Fig.~\ref{fig:shape} (top panel). Unfortunately, SPHERE data do not cover the northern hemisphere, so this part is based only on photometric data.

Next, we reconstructed the shape of Elektra from optical lightcurves and all 13 resolved images. The resulting shapes (middle and bottom panels of Fig.~\ref{fig:shape}) are smoother and lacking some of the features of the shape model based on SPHERE only resolved images. This is because of lower resolution of the Keck images that do not address any low-scale surface features. Specifically, the size of a pixel on the Keck II Elektra images corresponds to 12 to 26 km on Elektra's surface. Actually, it is $<$20 km for only one image. Considering these values and the Elektra's size of $\sim$200 km, the shape cannot be constrained to great details based on the Keck II images only. For instance, the bulge in the top-left panel of Fig.~\ref{fig:shape} is visible in the SPHERE only model. When we add the Keck images, the bulge in the model is mostly caused by shadowing. As the phase angle is 16 degrees, this is plausible.

All shape models based on different shape supports as well as on different amount of disk-resolved data are, in general, similar, and have dimensions and volume-equivalent sizes within only few percent. A volume-equivalent size based on various shape solutions is $D=199\pm7$~km, however, additional systematic uncertainties are difficult to reliably estimate. These mostly come from the uncertainty of the boundary condition in the AO data. Finally, sizes along Elektra's main axes are 262$\pm$7 x 205$\pm$6 x 164$\pm$5 km.

A comparison between all 15 deconvolved images and the corresponding model is shown in Fig.~\ref{fig:comparison}.

All shape models have pole orientations with ecliptic latitude close to $-$90$^{\circ}$, which is consistent with all previous determinations. The difference in the ecliptic longitude with respect to previous determinations is quite large, but this can be attributed to the fact that longitudes are dense for latitudes close to $\pm90^{\circ}$.

As an additional shape and size consistency check, we compared the 2D projections of our various shape models with the stellar occultation measurement in February 20, 2010. This observation \citep[already used in][]{Durech2011} consists of measurements from 8 stations spread along Elektra's shadow, including seven chords and one non-detection\footnote{\texttt{http://www.euraster.net/results/2010/20100220-Elektra-crd.gif}}. Because of the relatively large uncertainties in the timings of the chords and possible systematic offsets of individual chords, we decided not to include the occultation into the shape and size optimization, although such a procedure is supported in ADAM. Moreover, the occultation does not cover an interesting geometry, so it does not provide any useful additional information. All our scaled shape models of Elektra agree well with the occultation measurements, which demonstrates the reliability of our results.

\subsection{Bulk densities}\label{sec:densities}

The density of $1.60\pm0.13$ g\,cm$^{-3}$ was derived from our volume estimate and adopted mass of (6.6$\pm$0.4) 10$^{18}$ kg. Other estimates of Elektra's bulk density reported in \citet{Marchis2012a} ($1.7\pm0.3$ g\,cm$^{-3}$) and \citet{Hanus2013b} ($1.99\pm0.66$ g\,cm$^{-3}$) are consistent with our measurement within their large errors. This is the first bulk density of a triple asteroid obtained from ground-based observations that include disk-resolved images from the SHPERE instrument. The uncertainty of the bulk density is dominated by the mass uncertainty. The relative precision of the density is unusually high compared to typical values in the literature for other asteroids. This is due to the availability of the accurate mass estimate from the secondary moon orbit, combined with our accurate size estimate by the ADAM algorithm from the SPHERE resolved data.  

Other C-complex asteroids of similar size have a lower bulk density reported in the literature. \citet{Marchis2008,Marchis2008a} estimated the bulk density for (379) Huenna to $0.85\pm0.05$ g\,cm$^{-3}$ ($D\sim215$ km) and for (762) Pulcova to $0.9\pm0.1$ g\,cm$^{-3}$ ($D\sim140$ km). These measurements do not include an accurate estimate of the shape, since only the IRAS thermal infrared measurements were considered to estimate the size of the asteroids, so ignoring the existence of concavities, and an irregular shape. Consequently, those bulk densities should be considered as lower limits. With better shape models obtained combining AO, photometric and thermal observations, \citet{Descamps2009} reported a bulk-density of $1.4^{+0.5}_{-0.2}$ g\,cm$^{-3}$ ($D\sim187$ km) for the asteroid (121) Hermione, and \citet{Marchis2013a} a bulk density of $1.75\pm0.30$ g\,cm$^{-3}$ for (93) Minerva ($D\sim154$ km), and \citet{Marchis2012} a bulk density of $1.7\pm0.3$ g\,cm$^{-3}$ for (45) Eugenia in agreement with our measurement for Elektra. 


\section{Conclusions}\label{sec:conclusions}

We apply the ADAM shape modeling algorithm to an up-to-date optical disk-integrated dataset, 2 disk-resolved images obtained by the SPHERE instrument and 13 disk-resolved images from the Nirc2 of the Keck telescope, and derive the size and the first shape model of Elektra with local detail. The volume-equivalent diameter of $D=199\pm7$~km is currently the most reliable and precise size estimate of Elektra.

By combining the size estimate with the mass from \citet{Marchis2008a}, we computed Elektra's bulk density to be $\rho=1.60\pm0.13$ g\,cm$^{-3}$, which belongs to one of the most precise density determinations achieved so far for an asteroid.

Reliable bulk densities of other C-complex asteroids of similar size reported in the literature are usually consistent with the bulk density of Elektra.

\begin{acknowledgements}
JH greatly appreciates the CNES post-doctoral fellowship program. We are grateful to an anonymous referee for useful and constructive comments.

This research has made use of the Keck Observatory Archive (KOA), which is operated by the W. M. Keck Observatory and the NASA Exoplanet Science Institute (NExScI), under contract with the National Aeronautics and Space Administration.

\end{acknowledgements}

\bibliography{mybib}
\bibliographystyle{aa}

\input{tabs/tab1.tex}
\input{tabs/tab2.tex}

\input{tabs/tab3.tex}
\input{tabs/tab4.tex}

\end{document}

%% file: tabs/tab1.tex
\begin{table*}
\caption{\label{tab:spins}Rotational states of Elektra available in the literature as well as our new determination based on combined optical light curves and disk-resolved images from NIRC2 and SPHERE/IFS instruments mounted on W.M. Keck II and VLT/UT3 telescopes, respectively.}
\centering
\begin{tabular}{rrrr D{.}{.}{6} rr}
\hline 
 \multicolumn{1}{c} {$\lambda_1$} & \multicolumn{1}{c} {$\beta_1$} & \multicolumn{1}{c} {$\lambda_2$} & \multicolumn{1}{c} {$\beta_2$} & \multicolumn{1}{c} {$P$} & \multicolumn{1}{c} {Method} & \multicolumn{1}{c} {Original model} \\
 \multicolumn{1}{c} {[deg]} & [deg] & [deg] & [deg] & \multicolumn{1}{c} {[hours]} & \multicolumn{1}{c} {} & \multicolumn{1}{c} {published by} \\
\hline\hline
 190 & $-$81 &  &  & 5.22468 & E & \citet{Drummond1988b} \\
 180 & $-$85 & 240 & $-$40 & 5.22466 & E & \citet{Magnusson1990} \\
 190 & $-$81 &  &  & 5.224683 & E & \citet{Drummond1991} \\
 246 & $-$32 & 344 & $-$86 & 5.22466 & E & \citet{Michalowski1993} \\
 192 & $-$83 &  &  & 5.22468 & E & \citet{DeAngelis1995b} \\
 64 & $-$88 &  &  & 5.224664 & LI & \citet{Durech2007a}\\ 
 160 & $-$85 &  &  & 5.22466 & LI & \citet{Torppa2008} \\
 176 & $-$89 &  &  & 5.224663 & LI & \citet{Hanus2016a}\\ 
 64 & $-$90 &  &  & 5.224663 & ADAM & This work\\ 
 69 & $-$88 &  &  & 5.224663 & ADAM & This work\\ 
 71 & $-$88 &  &  & 5.224663 & ADAM & This work\\ 
 \hline
\end{tabular}
\tablefoot{
    The table gives ecliptic coordinates $\lambda$ and $\beta$ of all possible pole solutions, sidereal rotational period $P$, method used for the spin state determination (E -- methods assuming triaxial rotation ellipsoid shape models, LI -- lightcurve inversion with a convex shape approximation, ADAM -- shape model based on optical data and disk-resolved images), and reference to the corresponding publication. ADAM shape models of Elektra are reconstructed from disk-integrated optical data and (i)~raw SPHERE images (first), (ii)~all resolved images using subdivision surfaces shape support (second), and finally (iii)~all resolved data using octanoids shape support (last).
    }
\end{table*}

%% file: tabs/tab2.tex
\begin{table*}
\caption{\label{tab:photometry}List of optical disk-integrated light curves. For each light curve, the table gives the epoch, the number of points $N_p$, asteroid's distances to the Sun $r$ and Earth $\Delta$, used photometric filter and observation information.}
\centering
\begin{tabular}{rlr rr r rrr}
\hline 
\multicolumn{1}{c} {N} & \multicolumn{1}{c} {Epoch} & \multicolumn{1}{c} {$N_p$} & \multicolumn{1}{c} {$r$} & \multicolumn{1}{c} {$\Delta$} & \multicolumn{1}{c} {Filter} & Site & Observer  & Reference \\
 &  &  & [a.u.] & [a.u.] &  &  &  &  \\
\hline\hline
  1 & 1980-07-04.4 &  15 & 3.02 & 2.06 & V & TMO & Harris, Young & \citet{Harris1989a} \\
  2 & 1980-07-05.3 &  27 & 3.02 & 2.05 & V & TMO & Harris, Young & \citet{Harris1989a} \\
  3 & 1980-07-06.2 &  16 & 3.02 & 2.05 & V & TMO & Harris, Young & \citet{Harris1989a} \\
  4 & 1981-11-06.4 &  22 & 2.48 & 1.62 & V & KPNO & - & \citet{Weidenschilling1987} \\
  5 & 1981-12-02.3 &  19 & 2.51 & 1.68 & V & KPNO & - & \citet{Weidenschilling1987} \\
  6 & 1981-12-07.4 &   9 & 2.52 & 1.71 & V & TMO & Harris, Young & \citet{Harris1989a} \\
  7 & 1981-12-08.3 &  18 & 2.52 & 1.72 & V & TMO & Harris, Young & \citet{Harris1989a} \\
  8 & 1982-01-09.2 &  18 & 2.57 & 2.01 & V & KPNO & - & \citet{Weidenschilling1987} \\
  9 & 1982-01-14.2 &  10 & 2.58 & 2.06 & V & KPNO & - & \citet{Weidenschilling1987} \\
 10 & 1982-01-15.3 &   8 & 2.58 & 2.08 & V & KPNO & - & \citet{Weidenschilling1987} \\
 11 & 1982-12-16.4 &  17 & 3.27 & 2.79 & V & KPNO & - & \citet{Weidenschilling1987} \\
 12 & 1982-12-17.3 &  13 & 3.27 & 2.78 & V & KPNO & - & \citet{Weidenschilling1987} \\
 13 & 1983-03-23.3 &  28 & 3.45 & 2.65 & V & KPNO & - & \citet{Weidenschilling1987} \\
 14 & 1984-01-12.3 &  14 & 3.77 & 3.66 & V & KPNO & - & \citet{Weidenschilling1987} \\
 15 & 1984-01-15.3 &  10 & 3.77 & 3.61 & V & KPNO & - & \citet{Weidenschilling1987} \\
 16 & 1984-01-16.3 &  18 & 3.77 & 3.60 & V & KPNO & - & \citet{Weidenschilling1987} \\
 17 & 1984-04-10.3 &  31 & 3.79 & 2.86 & V & KPNO & - & \citet{Weidenschilling1987} \\
 18 & 1984-04-11.4 &  11 & 3.79 & 2.86 & V & KPNO & - & \citet{Weidenschilling1987} \\
 19 & 1984-07-05.2 &   6 & 3.78 & 3.68 & V & KPNO & - & \citet{Weidenschilling1987} \\
 20 & 1985-06-27.3 &  26 & 3.39 & 2.54 & V & KPNO & - & \citet{Weidenschilling1987} \\
 21 & 1986-06-13.2 &   9 & 2.66 & 2.33 & V & KPNO & - & \citet{Weidenschilling1987} \\
 22 & 1986-06-14.2 &   8 & 2.66 & 2.31 & V & KPNO & - & \citet{Weidenschilling1987} \\
 23 & 1986-06-15.2 &  11 & 2.66 & 2.30 & V & KPNO & - & \citet{Weidenschilling1987} \\
 24 & 1986-06-17.2 &   7 & 2.66 & 2.27 & V & KPNO & - & \citet{Weidenschilling1987} \\
 25 & 1988-01-01.3 &  17 & 2.93 & 2.01 & V & ESO, Chile & - & \citet{Debehogne1990} \\
 26 & 1988-01-02.2 &  25 & 2.93 & 2.01 & V & ESO, Chile & - & \citet{Debehogne1990} \\
 27 & 1988-01-03.2 &  31 & 2.94 & 2.01 & V & ESO, Chile & - & \citet{Debehogne1990} \\
 28 & 1988-01-04.2 &  35 & 2.94 & 2.01 & V & ESO, Chile & - & \citet{Debehogne1990} \\
 29 & 1991-06-14.2 &  31 & 3.04 & 2.20 & R & MCO & Danforth, Ratcliff & \citet{Danforth1994} \\
 30 & 1991-07-15.2 &  56 & 2.97 & 2.00 & R & MCO & Danforth, Ratcliff & \citet{Danforth1994} \\
 31 & 1994-02-11.8 &  34 & 3.40 & 2.41 & V & AOKU & - & \citet{Shevchenko1996} \\
 32 & 1994-02-13.8 &  18 & 3.40 & 2.41 & V & AOKU & - & \citet{Shevchenko1996} \\
 33 & 1994-02-15.0 &  24 & 3.40 & 2.41 & V & AOKU & - & \citet{Shevchenko1996} \\
 34 & 1994-02-15.7 &  24 & 3.40 & 2.42 & V & AOKU & - & \citet{Shevchenko1996} \\
 35 & 1994-02-17.8 &  25 & 3.41 & 2.42 & V & AOKU & - & \citet{Shevchenko1996} \\
 36 & 1994-03-18.8 &  45 & 3.46 & 2.61 & V & AOKU & - & \citet{Shevchenko1996} \\
 37 & 1994-05-15.9 &  11 & 3.55 & 3.45 & V & AOKU & - & \citet{Shevchenko1996} \\
 38 & 2001-04-22.9 &  40 & 3.70 & 2.82 & C & GO & Sposetti &     \citet{Durech2007a} \\
 39 & 2001-04-23.0 &  60 & 3.70 & 2.82 & C & GO & Sposetti &     \citet{Durech2007a} \\
 40 & 2003-10-21.9 & 557 & 2.50 & 1.70 & V & Craigie & Bolt &    \citet{Durech2007a} \\
 41 & 2003-10-23.9 & 595 & 2.50 & 1.69 & V & Craigie & Bolt &    \citet{Durech2007a} \\
 42 & 2003-10-24.9 & 614 & 2.51 & 1.69 & V & Craigie & Bolt &    \citet{Durech2007a} \\
 43 & 2003-10-29.9 & 626 & 2.51 & 1.68 & V & Craigie & Bolt &    \citet{Durech2007a} \\
 44 & 2003-11-14.6 &  61 & 2.53 & 1.66 & V & MTO & Bembrick &    \citet{Durech2007a} \\
 45 & 2003-11-15.7 &  72 & 2.53 & 1.66 & V & MTO & Bembrick &    \citet{Durech2007a} \\
 46 & 2003-11-19.6 &  83 & 2.53 & 1.67 & V & MTO & Bembrick &    \citet{Durech2007a} \\
 47 & 2003-11-27.7 & 112 & 2.54 & 1.69 & V & MTO & Bembrick &    \citet{Durech2007a} \\
 48 & 2003-11-29.7 & 116 & 2.55 & 1.70 & V & MTO & Bembrick &    \citet{Durech2007a} \\
 49 & 2009-12-12.1 & 144 & 2.88 & 2.07 & C & B81 & Salom, Esteban, Behrend & \citet{Hanus2016a} \\
 50 & 2011-03-07.9 & 102 & 3.66 & 2.69 & C & 615 & Montier, Behrend & \citet{Hanus2016a} \\
 51 & 2011-03-22.0 &  95 & 3.67 & 2.71 & C & 615 & Montier, Behrend & \citet{Hanus2016a} \\
 52 & 2011-03-23.9 & 110 & 3.67 & 2.72 & C & 615 & Montier, Behrend & \citet{Hanus2016a} \\
 53 & 2011-04-08.9 & 240 & 3.69 & 2.83 & C & C62 & Casalnuovo &  \citet{Hanus2016a} \\
 54 & 2011-04-09.9 & 242 & 3.69 & 2.84 & C & C62 & Casalnuovo, Chinaglia & \citet{Hanus2016a} \\
\hline
\end{tabular}
\tablefoot{
    TMO -- Table Mountain Observatory, CA, USA. KPNO -- Kitt Peak National Observatory. MCO -- Middlebury College Observatory. AOKU -- Astronomical Observatory of Kharkov University. GO -- Gnosca Observatory, Switzerland. MTO -- Mt Tarana Observatory, Bathurst, Australia. B81 -- Observatorio Astron\' omico Caimari. 615 -- Astroqueyras, Mairie, F-05350 Saint-V\' eran, France. C62 -- Eurac Observatory, Bolzano, Italy.
    }
\end{table*}

%% file: tabs/tab3.tex
\begin{table*}
\caption{\label{tab:ao}List of disk-resolved images. For each observation, the table gives the epoch, the telescope, the photometric filter, the exposure time, the airmass, R.A. and Dec of the asteroid, the distance to the Earth $\Delta$ and the reference or the PI of the project at Keck.}
\centering
\begin{tabular}{rrr rr rr rrr}
\hline 
\multicolumn{1}{c} {Date} & \multicolumn{1}{c} {UT} & \multicolumn{1}{c} {Instrument} & \multicolumn{1}{c} {Filter} & \multicolumn{1}{c} {Exp} & \multicolumn{1}{c} {Airmass} & \multicolumn{1}{c} {R.A.} & \multicolumn{1}{c} {Dec} & \multicolumn{1}{c} {$\Delta$} & Reference or PI \\
\hline\hline
 2002-09-22 &     07:11:29 &   Keck/NIRC2 &     H &   5.0 & 1.33 & 19 03 56 & -12 48 17 & 2.38 &               Dumas \\
 2002-09-22 &     07:50:29 &   Keck/NIRC2 &     K &   5.0 & 1.52 & 19 03 56 & -12 48 33 & 2.38 &               Dumas \\
 2002-09-22 &     07:53:34 &   Keck/NIRC2 &     H &   5.0 & 1.53 & 19 03 56 & -12 48 33 & 2.38 &               Dumas \\
 2002-09-27 &     07:15:13 &   Keck/NIRC2 &    Kp &   5.0 & 1.44 & 19 06 30 & -13 26 52 & 2.43 &             Merline \\
 2003-12-07 &     07:15:41 &   Keck/NIRC2 &    Kp &   3.0 & 1.43 & 03 45 25 & -15 58 22 & 1.74 & \citet{Marchis2008a}\\
 2005-01-15 &     12:25:31 &   Keck/NIRC2 &    Kp &   5.0 & 1.03 & 10 09 43 & 08 39 49 & 2.49 &  \citet{Marchis2008a}\\
 2005-01-15 &     14:14:01 &   Keck/NIRC2 &    Kp &   5.0 & 1.08 & 10 09 40 & 08 40 24 & 2.49 &  \citet{Marchis2008a}\\
 2008-06-06 &     14:31:34 &   Keck/NIRC2 &    Kp &  10.0 & 1.22 & 22 51 43 & -04 44 29 & 2.45 &         Engeneering \\
 2012-06-25 &     06:05:58 &   Keck/NIRC2 &    Kp &   7.0 & 1.03 & 14 45 27 & 12 44 46 & 3.08 &              Merline \\
 2012-07-14 &     08:23:15 &   Keck/NIRC2 &    Kp &   6.0 & 1.32 & 14 44 31 & 10 46 22 & 3.29 &           Armandroff \\
 2012-07-14 &     08:27:06 &   Keck/NIRC2 &     H &   6.0 & 1.34 & 14 44 31 & 10 46 21 & 3.29 &           Armandroff \\
 2012-08-10 &     06:00:58 &   Keck/NIRC2 &    Kp &   3.0 & 1.17 & 14 53 18 & 07 18 31 & 3.61 &              Merline \\
 2012-08-11 &     05:58:35 &   Keck/NIRC2 &    Kp &   3.0 & 1.18 & 14 53 50 & 07 10 21 & 3.63 &              Merline \\
 2014-12-09 &     01:37:51 &   VLT-UT3/SPHERE &   H & -- & 1.04 & 03 17 10 & -17 03 49 & 1.78 &   \citet{Yang2016} \\
 2014-12-30 &     01:03:02 &   VLT-UT3/SPHERE &   H & -- & 1.02 & 03 11 28 & -13 56 16 & 1.96 &   \citet{Yang2016} \\
\hline
\end{tabular}
\end{table*}

%% file: tabs/tab4.tex
\begin{table*}
\caption{\label{tab:densities}Sizes and densities of Elektra that are available in the literature as well as our new determination based on combined optical light curves and disk-resolved images from NIRC2 and SPHERE/IFS instruments mounted on W.M. Keck II and VLT/UT3 telescopes, respectively.}
\centering
\begin{tabular}{ccccc}
\hline 
 \multicolumn{1}{c} {$a$ x $b$ x $c$} & \multicolumn{1}{c} {$D_\mathrm{eq}$} & \multicolumn{1}{c} {$\rho_{\mathrm{bulk}}$} & \multicolumn{1}{c} {Method} & \multicolumn{1}{c} {Reference} \\
 \multicolumn{1}{c} {km} & \multicolumn{1}{c} {km} & \multicolumn{1}{c} {[g\,cm$^{-3}$]} & \multicolumn{1}{c} {} & \multicolumn{1}{c} {} \\
\hline\hline
  & 182$\pm$12 & & Thermal model from IRAS & \citet{Tedesco2004}\\ 
  & 191 & & Mean size from AO image & \citet{Marchis2006}\\ 
  & 215$\pm$15 & 1.3$\pm$0.3 & AO images from Keck, VLT and Gemini & \citet{Marchis2008a}\\ 
  & 183.0$\pm$2.3 & & Thermal model from AKARI & \citet{Usui2011}\\ 
  & 198.9$\pm$4.1 & & Thermal model from WISE & \citet{Masiero2011}\\ 
  & 191$\pm$14 & & Convex shape + occult & \citet{Durech2011}\\ 
  & 174.9$\pm$25.5 & 2.34$\pm$0.34 & Thermal model of Spitzer spectra & \citet{Marchis2012b}\\ 
  & 197$\pm$20 & 1.6$\pm$0.5 & Thermophysical model of Spitzer spectra & \citet{Marchis2012b}\\ 
  & 189.62$\pm$6.81 & 1.84$\pm$0.22 & Compilation & \citet{Carry2012b}\\ 
  & 161.94$\pm$3.82 & & Thermal model from WISE 3band  & \citet{Masiero2012}\\ 
  & 185$\pm$20 & 1.99$\pm$0.66 & Convex shape + Keck AO & \citet{Hanus2013b}\\ 
  258x203x163 & 196$\pm$5 & & ADAM: LCs + SPHERE, subdivision & This work\\ 
  263x204x165 & 200$\pm$5 & & ADAM: LCs + all AO, subdivision & This work\\ 
  265x208x163 & 201$\pm$5 & & ADAM: LCs + all AO, octanoids & This work\\ 
  262x205x164 & 199$\pm$7 & $1.60\pm0.13$ & ADAM: multiple models & This work\\ 
 \hline
\end{tabular}
\tablefoot{
    The table gives dimensions along the three main axis, the volume-equivalent diameter $D_\mathrm{eq}$, the method/dataset used for the spin state determination, and the reference to the corresponding publication. ADAM shape models of Elektra are reconstructed from disk-integrated optical data and (i)~raw SPHERE images (first), (ii)~all resolved images using subdivision surfaces shape support (second), and finally (iii)~all resolved data using octanoids shape support (last). The bulk density estimate is assuming mass of 6.6$\pm$0.4 10$^{18}$~kg from \citet{Marchis2008a}.
    }
\end{table*}